\documentclass[smallcondensed,natbib]{svjour3}
\usepackage[utf8]{inputenc}
\usepackage{booktabs}
\usepackage{graphicx}
\usepackage{url}
\usepackage{amsmath}
\usepackage{subfig}
\usepackage{listings}
\usepackage{threeparttable}

\title{Large-scale Information Retrieval in Software Engineering - An 
Experience Report from Industrial Application}
\author{Michael Unterkalmsteiner, Tony Gorschek, Robert Feldt and Niklas 
Lavesson}

\institute{M. Unterkalmsteiner, T. Gorschek, R. Feldt \at Department of 
Software Engineering, Blekinge Institute of Technology 
\email{\{mun,tgo,rfd\}@bth.se}
\and
N. Lavesson \at Department of Computer Science and Engineering, Blekinge 
Institute of Technology \email{nla@bth.se}
}

\begin{document}

\maketitle

\begin{abstract}
\emph{Background:} Software Engineering activities are information intensive. 
Research proposes Information Retrieval (IR) techniques to support engineers in 
their daily tasks, such as establishing and maintaining traceability links, 
fault identification, and software maintenance.
\emph{Objective:} We describe an engineering task, test case selection, and 
illustrate our problem analysis and solution discovery process. The objective 
of the study is to gain an understanding of to what extent IR techniques (one 
potential solution) can be applied to test case selection and provide decision 
support in a large-scale, industrial setting.
\emph{Method:} We analyze, in the context of the studied company, how test case 
selection is performed and design a series of experiments evaluating the 
performance of different IR techniques. Each experiment provides lessons learned
from implementation, execution, and results, feeding to its successor.
\emph{Results:} The three experiments led to the following observations: 1) 
there is a lack of research on scalable parameter optimization of IR techniques 
for software engineering problems; 2) scaling IR techniques to industry data is 
challenging, in particular for latent semantic analysis; 3) the IR context 
poses constraints on the empirical evaluation of IR techniques, requiring more 
research on developing valid statistical approaches.
\emph{Conclusions:} We believe that our experiences in conducting a series of 
IR experiments with industry grade data are valuable for peer researchers so 
that they can avoid the pitfalls that we have encountered. Furthermore, we 
identified challenges that need to be addressed in order to bridge the gap 
between laboratory IR experiments and real applications of IR in the industry.
\end{abstract}

\section{Introduction}
The field of Software Engineering thrives on the continuous interchange of 
knowledge and experience between those who research and those who 
practice~\citep{wohlin_success_2012}. 
Research without application is, in the long-term, meaningless and application 
without research leads to stagnation~\citep{ivarsson_method_2011}. This view 
requires that research efforts are seeded by concrete problem statements from 
industry, and that solutions are evaluated within industrial applications and 
validated by practitioners~\citep{gorschek_model_2006}. Scientific advances 
require also that results are reproducible, which in turn is only possible with 
disciplined reporting that covers the necessary details to replicate studies. 
\cite{shepperd_researcher_2014} performed a meta-analysis on 600 
software defect prediction (SDP) results published in literature and determined 
that differences in prediction performance varies with the research group 
rather than the studied SDP technique. They suggest that this researcher bias 
could be addressed by improving the communication of study designs and details 
of the used technologies. 

On that note, even though we address a different issue than SDP, we report on 
experiences in identifying, developing and evaluating a potential solution for 
our industrial partner: test case selection~\citep{rothermel_analyzing_1996} 
decision support by Information Retrieval 
techniques~\citep{grossman_information_2004}. The main goal of this paper is to 
illustrate the path from problem identification to solution definition and 
evaluation, and to dissect the taken design decisions that address a real-world 
software engineering problem. With complex problem statements it is sensible to 
develop new solutions upon existing and validated building blocks. We 
identified IR techniques, as they are applied in traceability 
recovery~\citep{borg_recovering_2013} in general 
and feature location~\citep{dit_feature_2011} in particular, as part of a 
potential solution and performed a series of experiments with the goal to 
evaluate whether the overall approach is feasible. We illustrate the design and 
evolution of the experiments, pointing out practical implementation challenges 
that we encountered while adapting IR techniques proposed by research. Even 
though this research was driven by eventually developing a workable solution 
for test case selection decision support, we focus in this paper on the lessons 
learned from designing and executing the experiments. Thereby, we make the 
following contributions:

\begin{itemize}
 \item An application of IR techniques to support test case selection in a 
 Software Product Line context.
 \item An illustration of three experimental setups, each leading to a set of 
new research challenges for IR applications on industry-grade data.
\end{itemize}

The remainder of this paper is structured as follows. In 
Section~\ref{sec:BackRel} we present the context in which we conducted the 
research, illustrate the envisioned solution and discuss related work. 
Section~\ref{sec:problem_definition} provides the formal problem definition 
that guides the overall design of the experiment. In Section~\ref{sec:expsetup} 
we illustrate the design, execution, results and lessons learned from each of 
the three experimental setups. The paper concludes in 
Section~\ref{sec:conclusion}, pointing out avenues for future work.

\section{Background and related work}\label{sec:BackRel}
This research is conducted in collaboration with ES (``Embedded Systems'', name 
anonymized for confidentiality reasons) and the solution development is driven 
by their particular context and requirements. In this section, we first 
give an overview of the case context and problem 
(Section~\ref{sub:casecontext}). We take then a step back and analyze the 
problem with respect to the state of art (Section~\ref{sub:soa}) before 
sketching a solution in Section~\ref{sub:idea}. We discuss related work in 
Section~\ref{sub:related_work}.

\subsection{Case context}\label{sub:casecontext}
ES, developing both hard- and software for their worldwide marketed products,
has a three-decade history in developing embedded systems, although the 
particular applications areas have changed over time. 

\subsubsection{Variability management and consequences for quality 
assurance}\label{sub:vmqa}
In autumn 2011, we performed a lightweight process 
assessment~\citep{pettersson_practitioners_2008} at ES in order to identify 
improvement opportunities, in particular in the coordination between 
requirements engineering and software testing. We interviewed 16 employees 
(line managers, product and project managers, test and technology leads, and 
test engineers) and reviewed project documentation, test runs and trouble 
reports of two recently completed projects. As observed by 
\cite{thorn_current_2010} in small and medium sized companies, we identified 
also at ES a lack of variability management in the problem space (requirements 
analysis, specification and maintenance). This has historical reasons, mostly 
attributed to the strong technical support for managing variants in the 
solution space (separate development of a platform and product specific 
software). ES generally develops a new product generation on the 
basis of their current products, reusing and extending the existing 
requirements specifications. This has the advantage that the requirements 
management process is lightweight and requires minimal documentation. However, 
this can also lead to challenges for impact analysis of new features since 
relationships between requirements are not 
documented~\citep{graaf_embedded_2003}. The interviewed test engineers at ES 
also indicated that without variability management, they may select system test 
cases that are not applicable for a particular product. This can lead to 
a lower efficiency in test execution as information on applicability needs to 
be retrieved from the project (project manager or product expert) or by an 
in-depth analysis of the particular test case and product. Looking at the test 
case database at ES, for a typical product release, 400-1000 system test-cases 
are selected, run and evaluated, each consisting of up to 20 manual test steps. 
An improved test case selection can therefore reduce rework and time-to-market.

ES initiated different programs to improve their quality assurance efficiency. 
They introduced a risk-based strategy in order to focus the testing effort on 
those parts of the product carrying a risk to contain faults, optimizing 
thereby available time and resources. The risk-based test selection is based 
upon expert judgments from technology leads, product experts and test 
maintainers, but also on historical data from test runs of similar products. 
Furthermore, automated integration tests are run on every version control 
system commit and before the product is handed over to the quality assurance 
(QA) department. A third avenue to achieve a more precise 
test case selection, developed at ES, is described next.

% 
% System test-cases are not automated and consist of up to 20 manual test 
% steps a QA engineer has to perform in order to verify that the product under 
% test fulfills the stated acceptance criteria. For a product release, 
%400-1,000 
% system test cases are selected, run and evaluated. The selection of 
% system test cases is based on historical test runs. Functionality that 
% has been added to a new product is tested by newly created test-cases. QA 
% engineers recognize that the lack of explicit support for product variation 
% management causes issues that make it difficult to provide system tests on 
%time 
% and with the required quality:
% \begin{itemize}
%  \item Functionality may be delivered untested, since there is no 
% commonly used, reliable, and up-to-date repository that specifies the 
% functionality of a product before it is delivered to QA. 
%  \item Test-cases may be selected, although the functionality is not present 
% in the tested product, leading to delays in the test execution. QA needs to 
% reassure that the functionality in question is indeed excluded, querying 
% developers, project managers or product experts.
% \end{itemize}
% 
% Axis Communications AB realized, in the light of a growing product portfolio, 
% the importance of these challenges. Therefore, they proactively initiated 
% several improvement efforts, one of which is to identify means for test case 
% selection improvement.

\subsubsection{Test case selection based on product 
configurations}\label{sub:sop}
For ES, an important criterion to adopt any test case selection 
support is to reuse as many existing resources and artifacts as possible, 
causing very little additional up-front investments to the overall development 
process.
The developed approach leverages on the existing technical 
infrastructure managing product variants in the solution space, where a central 
configuration file determines the activation state of a feature in the product. 
The approach consists of:
\begin{itemize}
	\item a parser that determines the activation state of a feature from the 
	central configuration file,
	\item the creation of a mapping between features and test-case modules 
	(containing a set of related test-cases), verified by product experts.
\end{itemize}

A product delivered to QA can be analyzed automatically to 
determine the activation status of its features. The parser creates a Product 
Configuration Report (PCR) that is used by QA engineers to guide the 
selection of system test cases. Testers need thereby to map between product 
features and test modules. In the current implementation, this solution has the 
following weak points:
\begin{itemize}
	\item The feature list in the PCR is incomplete, i.e. some 
	test case modules can not be mapped to a feature.
	\item Some test-cases need to be executed for every product. These test 
	cases 
	are also not mapped to a feature in the PCR.
	\item The granularity of the feature list in the PCR disallows a mapping 
	to specific test cases. 
	\item The mapping is static. If the organization of the test case 
	modules changes or features in the central configuration file change, 
	testers 
	and product experts need to re-verify the mapping.
\end{itemize}

Based on these observations we started to explore alternative approaches to 
address the test case selection problem.

\subsection{Overview of the state of art}\label{sub:soa}
As illustrated in Section~\ref{sub:vmqa}, ES resides in a Software Product Line 
(SPL)~\citep{clements_software_2001} context, where verification and validation 
is difficult due to the large number of product combinations that can be 
created~\citep{perrouin_automated_2010}. This problem 
is aggravated when the product is a combination of software and hardware, since 
late identification of defects when verifying an embedded system is 
costly~\citep{broy_challenges_2006,ebert_embedded_2009}. Model-based testing 
(MBT) techniques~\citep{utting_taxonomy_2012} have been proposed to address 
this complex challenge from an SPL perspective (see 
\cite{engstrom_software_2011} for a review of solution proposals). However, 
many of these proposals assume that companies apply variability 
management~\citep{babar_managing_2010} in their 
product lines, allowing the derivation of test strategies and the efficient 
selection of test cases. \cite{chen_systematic_2011} reviewed the variability 
management techniques proposed in literature, showing that the majority is not 
validated in an industrial context. This is supported by the observation by 
\cite{thorn_current_2010} that companies have control over variation in the 
solution space, lack however techniques and methods for variability 
management in the product and problem space. With a lack of variability 
management in the problem space, as it is the case for our case 
company, software product line testing principles~\citep{pohl_software_2006} or 
test case selection techniques~\citep{lee_survey_2012} are not applicable 
without upfront investment. Since reuse of existing resources and artifacts is 
a major criterion for ES to adapt a solution (see Section~\ref{sub:sop}), we 
sought for alternatives.

Test case selection, in the context of regression testing, encompasses the 
problem of choosing a subset of test cases, relevant to a change in the system 
under test, from a set of existing test cases~\citep{rothermel_analyzing_1996}. 
To address this engineering problem, various techniques have been developed 
(see \cite{engstrom_systematic_2010} and \cite{yoo_regression_2012} for 
comprehensive reviews). In general, they can be classified into white and 
black-box selection techniques. The former rely upon structural knowledge of 
the system under test, exploiting, for example, changes in source code, data 
and control flow, or execution traces~\citep{yoo_regression_2012}. The latter 
use other work products, such as design documentation, to perform impact 
analysis~\citep{yoo_regression_2012}. Both white and black-box test 
case selection techniques assume some sort of traceability between the changed 
artifact and the relevant test cases. Since the lack of variability management 
excludes the black-box selection approaches, ES has implemented traceability 
between product configurations and test cases as described in 
Section~\ref{sub:sop}, with the observed drawbacks. Next, we give a motivation  
and outline for the solution we aim to implement and evaluate in the remainder 
of this paper.

\subsection{Solution development}\label{sub:idea}
\begin{table}
 \caption{Candidate artifacts for automated feature 
identification}\label{tab:candidates}
 \centering
 \begin{tabular}{p{3.5cm}p{3.5cm}p{3.5cm}}
 \toprule
 \emph{Artifact} & \emph{Benefits} & \emph{Liabilities} \\
 \midrule
 \emph{PFD} is a design document written in natural language that details of 
what a certain functionality consists of and 
how it is intended to be used. & Detailed and accurately linked to a particular 
product variant. & Exists only for newly introduced features, i.e. is
incomplete w.r.t. the overall functionality of a product variant. \\
\midrule
 \emph{PCS} contains all available configuration 
options and dependencies among them, used to configure a product at build-time 
& Straightforward identification of activated/de-activated features & Not 
all tested features are activated/de-activated at build-time \\
\midrule
 \emph{Source code} & Complete w.r.t. identifying both build-time and 
run-time bound functionality. & Lowest possible abstraction level for 
representing features \\
\midrule
 \emph{Version Control} data & Provides semantic information to source code 
changes, such as commit comments, time, scale and frequency of changes. & 
Except for commit comments, little use for feature identification if not used 
in combination with source code. \\ 
 \bottomrule
 \end{tabular}
\end{table}
We outline the solution development approach by breaking down the test case 
selection problem into two objectives.

The first objective is to identify the to-be-tested features in a 
particular product variant that is delivered to QA. Currently, product variants 
are not managed in the problem space (e.g. in the requirements specifications) 
at ES. Hence, other work products that are created during the project and that 
are up-to-date when the product is delivered to QA need to be used instead. 
Table~\ref{tab:candidates} summarizes those candidate artifacts 
that fulfill these criteria.

The second objective is to map system test-cases to identified features. A 
system test-case contains information on the overall objectives of the 
test, acceptance criteria, and test steps. By establishing a trace link between 
a feature in a particular product variant and the set of relevant test-cases 
for that feature, test-case selection can be performed. In ES's solution, the 
mapping between the PCR and test-case modules is performed manually, which 
leads to the drawbacks illustrated in Section~\ref{sub:sop}.

Looking at the available artifacts in Table~\ref{tab:candidates}, only 
the source code artifacts provide complete information on the functionality in 
a particular product variant. On the other hand, the source code information 
(comments, method and variable identifiers, and literals) regarding each
feature is given at a low abstraction level, compared to the other work 
products, i.e. the PCS and the PFD.

Inspired by ES's semi-automatic solution that uses the PCS and expert judgment 
for test-case selection, we envision a solution that is applicable on source 
code, providing feature existence information, and is not dependent on regular 
expert input. In the area of source code mining, in particular textual feature 
location techniques based on natural language 
processing (NLP)\footnote{Note that the terms NLP and IR are independently used 
in literature to describe computerized processing of text and there is an 
overlap on what they comprise~\citep{falessi_empirical_2013}. In the 
remainder of the paper we use exclusively the term IR to maintain 
consistency.} or information retrieval (IR)~\citep{grossman_information_2004} 
seem promising. The goal of feature 
location is to support software developers in maintenance tasks by identifying 
relevant locations in the source code, e.g. to remove a fault or to extend a 
feature. The premise of textual feature location is that comments, identifiers, 
and literals encode domain knowledge, representing features than can be located 
in the source code by a set of similar terms~\citep{dit_feature_2011}. 

The test case selection problem and the two outlined objectives at the 
beginning of this subsection can be formulated as a feature location problem:
Given a test-case, representing a particular feature, locate the source code 
that implements that feature in the given product variant. The test case serves 
thereby as query, returning a list of artifacts ranked according to their 
similarity to the test case. Note that the specific problem of feature location 
can be also seen in the wider context of traceability recovery. The decision 
whether the given test case is 
selected or not, depends on whether the highest ranked artifact has a 
similarity score above or below a threshold $\alpha$. The particular value for 
$\alpha$ is not known in advance and must be determined experimentally, with a 
confidence interval such that the test engineer can gauge the robustness of the 
suggestion.  Determining $\alpha$ can be achieved by sampling a set of features 
for which the following is known: (a) the test case(s) verifying the feature 
and (b) the source code implementing the feature. We call this relationship 
between feature, test case(s) and source code a feature chain. Once the value 
and confidence interval for the threshold $\alpha$ is determined, it can be 
used to perform test case selection. 

Note that this solution description is the starting point and not the outcome 
of our investigation. We provide a more formal problem definition in 
Section~\ref{sec:problem_definition} and illustrate three experimental setups 
(Section~\ref{sec:expsetup}) that were designed to study to what degree IR 
techniques can be used to support test case selection with industry grade data.

\subsection{Related work}\label{sub:related_work}
%``Feature Location in a Collection of Product Variants'': see Section 4.D 
%which is basically the same way how we establish the ground truth. Section 5.A 
%introduces evaluation measures!

\subsubsection{Traceability Recovery}\label{sub:tr}
Mapping product features from source code to the respective test cases, as
described in Section~\ref{sub:idea}, is related to 
traceability~\citep{gotel_analysis_1994} in general, and to the concept 
assignment problem~\citep{biggerstaff_concept_1993} in particular. Concept 
assignment is the process of mapping concepts from the problem domain (e.g. a 
requirement or feature expressed in potentially ambiguous and imprecise natural 
language) to the solution domain (e.g. an algorithm expressed in precise and 
unambiguous numerical computations). Recovering traceability links between 
source code and natural language documentation using IR techniques was 
pioneered by \cite{antoniol_recovering_1999,antoniol_information_2000}, 
\cite{maletic_automatic_1999} and 
\cite{maletic_using_2000,maletic_supporting_2001}. These early studies 
envisioned the potential of IR techniques to support software engineers in 
program comprehension~\citep{maletic_automatic_1999,antoniol_recovering_1999}, 
requirement tracing and impact analysis, software reuse and 
maintenance~\citep{antoniol_recovering_1999}. 
Following these initial investigations, comprehensive experiments were 
conducted, studying particular IR techniques in depth (e.g. 
\cite{marcus_information_2004,zhao_sniafl:_2006,de_lucia_recovering_2007,
poshyvanyk_concept_2012}) or comparing newly proposed techniques to previously 
studied ones (e.g. \cite{marcus_recovering_2003,poshyvanyk_combining_2006, 
cleary_empirical_2009}). A large part of these studies use small scale data 
sets, derived from open source applications or public repositories. 
Furthermore, while a large body of work w.r.t. traceability recovery has been 
reviewed by \cite{borg_recovering_2013}, only a few addressed the recovery of 
links between source code and test cases (e.g. 
\cite{de_lucia_can_2006,de_lucia_assessing_2009, 
de_lucia_improving_2011}), and have been evaluated only on student 
projects~\citep{borg_recovering_2013}. 

Besides the study of individual IR techniques and their traceability recovery 
performance, investigations into the hybrid techniques show promise. 
\cite{gethers_integrating_2011} combined deterministic and probabilistic IR 
techniques, exploiting the fact that they capture different information. While 
this hybrid traceability recovery approach outperformed the individual 
techniques, determining the number of topics in the probabilistic 
approach~\citep{steyvers_probabilistic_2007} and finding an optimal 
combination of the individual approaches are still research 
challenges~\citep{gethers_integrating_2011}. More recently, 
\cite{qusef_recovering_2014} combined textual and runtime
information to recover traceability between source code and unit test cases. In 
their approach, first the runtime information is used to create a candidate 
list of traces, which is then further refined in a second step which analyzes 
textual information. While the evaluation of this hybrid technique indicates 
that it outperforms the precision of individual 
techniques~\citep{qusef_recovering_2014}, it requires the collection of a test 
execution trace and has been designed for unit-testing, i.e. white-box testing.

\subsubsection{Test case prioritization}\label{sub:tcp}
While test case selection encompasses the problem of identifying test cases 
relevant to a change in the system under test, prioritization aims at ordering 
test cases such that defects are detected as early as 
possible~\citep{yoo_regression_2012}. \cite{islam_multi-objective_2012} recover 
traceability links between source code and system requirements with IR 
techniques and use this information together with code coverage and test 
execution cost to identify optimal test orderings. This approach requires 
structural knowledge of the system under test, i.e. source code in this 
particular case. \cite{thomas_static_2014} propose therefore an approach that 
relies only upon data originating from the test cases. They use an IR technique 
to measure the similarity between test cases and prioritize them such that 
dissimilar test cases are run first.

\subsubsection{Configuration of IR techniques}\label{sub:config}
Numerous studies proposed and evaluated IR techniques for different software 
engineering tasks (see the systematic literature reviews by 
\cite{dit_feature_2011} and \cite{borg_recovering_2013}), leading to the need 
of systematically comparing and evaluating these techniques. The fundamental 
idea behind any configuration optimization is to guide the process by a 
data-driven parameter selection. For example, \cite{falessi_empirical_2013} use 
IR techniques to detect equivalent requirements and evaluate the performance of 
242 configurations. While they identify an optimal technique, they point out 
that the particular parameters are dependent on the underlying data set, which 
means that the identification of the optimal technique configuration requires 
considerable effort. \cite{lohar_improving_2013} propose therefore to use 
Genetic Algorithms~\citep{goldberg_genetic_1989} to search for the optimal IR 
technique configuration. A similar approach was proposed by 
\cite{panichella_how_2013} to configure probabilistic IR techniques (Latent 
Dirichlet Allocation~\citep{blei_latent_2003}).

%\subsubsection{Others}
%Meghan Revelle, Malcom Gethers, Denys Poshyvanyk: Using structural and textual 
%information to capture feature coupling in object-oriented software. Empirical 
%Software Engineering 16(6): 773-811 (2011)
%
%Bogdan Dit, Meghan Revelle, Denys Poshyvanyk: Integrating information 
%retrieval, execution and link analysis algorithms to improve feature location 
%in software. Empirical Software Engineering 18(2): 277-309 (2013)
%
%Tathagata Dasgupta, Mark Grechanik, Evan Moritz, Bogdan Dit, Denys Poshyvanyk: 
%Enhancing Software Traceability by Automatically Expanding Corpora with 
%Relevant Documentation. ICSM 2013: 320-329
%
%Bogdan Dit, Evan Moritz, Mario Linares Vásquez, Denys Poshyvanyk: Supporting 
%and Accelerating Reproducible Research in Software Maintenance Using TraceLab 
%Component Library. ICSM 2013: 330-339

\subsubsection{Contribution}
Traceability recovery supported by IR techniques has seen a lot of research in 
many application areas in software engineering (see Section~\ref{sub:tr}). 
However, research on traceability recovery with IR techniques between source 
code and test cases has been rare~\citep{borg_recovering_2013}, or not 
designed for black-box system testing (e.g. \cite{qusef_recovering_2014}).
In this paper we propose to use IR techniques for test case selection in the 
context of Software Product Lines. While IR techniques have been applied for 
test case prioritization (see Section~\ref{sub:tcp}), their use in test case 
selection has received less attention. IR techniques need to be 
configured in order to reach optimal performance. Recent studies show that 
configuring IR techniques should be driven by experiments since the performance 
depends on the underlying data (see Section~\ref{sub:config}). We use an 
industrial data set that is at least by an order of magnitude larger than in 
previous studies attempting to configure IR techniques. We report on the 
practical implementation of our proposed test selection techniques, the 
challenges we encountered and the lessons learned in evaluating the approach 
experimentally.

% To the best of our knowledge, IR techniques have not yet been used in the 
% context of test case selection. Furthermore, the performance of feature 
% location based on IR techniques has not been evaluated beyond preliminary, 
% proof-of-concept, applications~\citep{dit_feature_2011}. Therefore, the main 
% purpose of this study is to investigate the fitness of IR techniques for the 
% task of test-case selection by performing a set of experiments on industry-scale 
% data. 
% Our intuition is that the information contained in system test cases encodes 
% the to-be-tested feature and it can therefore be used to locate that feature 
% in the source code, by means of IR techniques based on textual similarity.
% In addition, we illustrate on a detailed level the encountered technical issues 
% both in terms of the data at hand and the technologies that were used to 
% implement IR in practice.

\section{Research method}\label{sec:problem_definition}
In Section~\ref{sub:idea} we stated the informal hypothesis that textual 
feature location based on IR techniques can be used to support test case 
selection decisions. Therefore, using the goal-question-metric 
method~\citep{basili_improve_1995} to specify the 
goal~\citep{dyba_evidence-based_2005} of this study, we:

\emph{Analyze IR techniques, for the purpose of evaluation, with respect to 
their support for test case selection, from the point of view of 
the practitioner, in the context of industry-grade software development 
artifacts.}

Based on this goal definition, we state the following research questions.
\begin{description}
 \item RQ-1 To what extent can state-of-the-art IR techniques be applied in a 
large-scale industrial context? 
 \item RQ-2 To what extent do the used software development artifacts influence 
the performance of IR techniques?
 \item RQ-3 To what extent can the studied IR techniques support test case 
selection?
\end{description}

We defined the problem in the context of our industrial partner and use their 
data to design, implement and evaluate a solution candidate. This approach 
differs from previous IR experiments on textual feature location which are 
mostly limited to preliminary evaluations in academic 
contexts~\citep{dit_feature_2011}. Hence, the purpose of RQ1 is 
to understand the challenges and to identify solutions for conducting IR 
experiments on data sets whose characteristics correspond to data sets from the 
problem domain. We conduct an experiment that evolves over three setups and 
address RQ-1 by reporting on lessons learned and analyzing them w.r.t. previous 
IR experiments reported in literature.
With RQ-2 we aim to understand the impact of input data characteristics on the 
performance of IR techniques. With IR performance we refer to both the 
techniques' efficiency in terms of computational cost and to the techniques' 
effectiveness (the specific effectiveness measures differ in the experimental 
setups and are defined there). We address RQ-2 in Experimental Setup 2 and 3. 
The purpose of RQ-3 is to compare the studied IR techniques and to determine 
whether the approach of using feature location as an aid to test case selection 
is feasible. We address RQ-3 in Experimental Setup 3.
The remainder of this section describes the overall design of our experiment.

\subsection{Description of objects}\label{sub:object_description}
The objects of the experiment are software development artifacts belonging to 
the implementation (source code, configuration files, user help documentation) 
and the system test (natural language test steps) domain. 

In the implementation domain, there is a set of features 
$F=\{F_1,\ldots,F_{nf}\}$, where $nf$ is the total number of features, that 
represent the configurable functionality of a family of systems $V$. The feature 
vector $A_i=[a_1,\ldots,a_{nf}]$, where $a_j=\{0,1\}$, identifies a particular 
variant $V(A_i)$ of the system family $V$. Furthermore, there is a set of source 
code files $C=\{C_1,\ldots,C_{nc}\}$, where $nc$ is the total number of 
source code files, each consisting of a set of source code lines 
$L_{Cj}=\{L_1,\ldots,L_{nl}\}$, where $nl$ is the total number of lines in 
that source code file. Depending on the particular implementation of a 
feature, there exists a mapping of feature $F_i$ to either: 
\begin{itemize}
  \item $C(F_i) \subset C$, i.e. a subset of the source code files
  \item $L_{Cj}(F_i) \subset L_{Cj}$, i.e. a subset of the source code lines
  \item a combination of the above, i.e. feature $F_i$ is mapped to a subset of 
  the source code files \emph{and} a subset of source code lines
\end{itemize}
Note that $C(F) \neq C$ and $L_{Cj}(F) \neq L_{Cj}$, i.e. there is a subset of 
source code files/lines that are not mapped to any configurable feature. Those 
source code files/lines are common to every system variant in $V$.

In the system test domain, there is a set of test cases 
$T=\{T_1,\ldots,T_{nt}\}$, where $nt$ is the total number of test cases. A 
test case is a natural language document that contains categorization 
information (test case name, test module name, test module information) and 
test execution information (assumptions, initialization information, acceptance 
criteria, test objective, test steps).

\subsection{Feature chains}
\begin{figure}
 \begin{center}
  \includegraphics[scale=0.5]{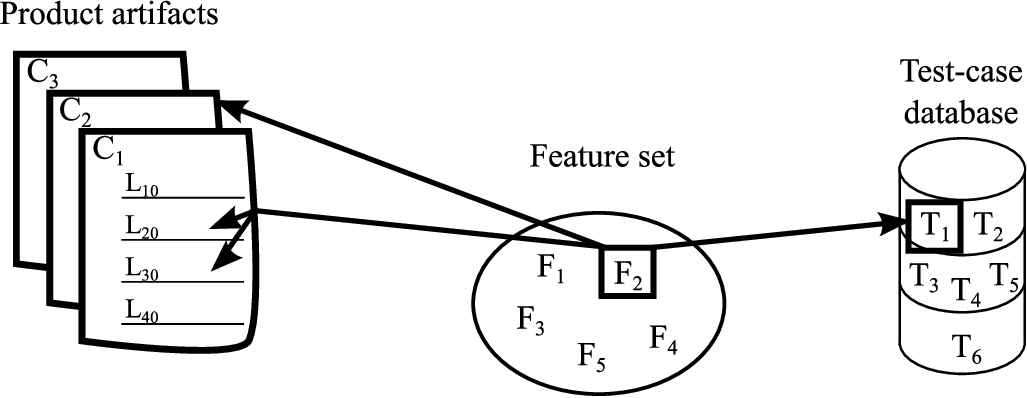}
 \end{center}
\caption{Example of a chain connecting a feature with test case and product 
artifacts}
\label{fig:chain} 
\end{figure}

Specific instances of the objects defined in 
Section~\ref{sub:object_description} determine a \emph{feature chain}, 
connecting features, source code and test cases (Figure~\ref{fig:chain}). A 
feature chain consists of two links:
\begin{enumerate}
 \item Feature to test cases: established by senior test engineers, linking a 
subset of all features $F$ to a subset of all test-cases $T$, i.e. $T(F_{1..n} 
\subset F) \subset T$, where $F_i \rightarrow T_{1..m}$.
 \item Feature to product artifacts: established by identifying the impact of 
the configuration switch activating/deactivating a feature on product artifacts 
(i.e. file and/or line addition/removal).
\end{enumerate}

A feature chain $I_i=\{F_i, C(F_i), L_{Cj}(F_i), T(F_i)\}$ connects the 
artifacts from the implementation and system test domain. 
Figure~\ref{fig:chain} illustrates this concept. Feature $F_2$ is verified by 
test case $T_1$. This connection is established by a senior test engineer. 
Feature $F_2$ is implemented in source code file $C_2$ and in lines $L_{20}$ to 
$L_{30}$ in file $C_1$. This connection is established by identifying the impact 
of the configuration switch activating/deactivating feature $F_2$ on $C_1$, 
$C_2$, and $C_3$.

\subsection{Independent variables}\label{sub:indepvar}
The potential independent variables, i.e. the variables that can be controlled 
and manipulated in the experiment, are numerous. Since the overall architecture 
of IR techniques is that of a pipeline, individual components can generally be 
replaced as long as input/output constrains are satisfied. 
\cite{falessi_empirical_2013} propose a classification of IR techniques 
consisting of four main components: IR model, term extraction, weighting 
schema, and similarity metric\footnote{\cite{grossman_information_2004} 
provides a broader overview}. For each of these dimensions, it is possible to 
identify multiple factors that deserve consideration in an experiment 
evaluating IR techniques.
However, since the main purpose of this study is to determine the 
feasibility of IR techniques for test case selection, we limit the number of 
independent variables and values to a realizable number 
(see Table~\ref{tab:iv}). As it turned out during the execution of the 
experiment, with increasing complexity and amount of data on which the IR 
techniques are applied, we had to remove values as they were not feasible to 
manipulate, but also \emph{added} values as the data required e.g. 
additional pre-processing. These decisions are motivated and documented in the 
corresponding experimental setup in Section~\ref{sec:expsetup}. We describe now 
the independent variables shown in Table~\ref{tab:iv} in more detail.

% [QUESTION: I list now in Table 3 the initially envisioned variants of NLP 
% techniques for the experiment. As we know now, it is illusory to run such an 
% experiment with actual industry data (time-wise and me not being capable to 
% implement everything since out-of-the-box tools are not available)... So what 
% to do: a) keep the variations here in the experiment setup an explain later 
% that they could not be implemented or b) not list the variants --> but then we 
% have no experiment...]

\begin{table}
 \caption{Selected independent variables (IV) and values}\label{tab:iv}
 \centering
 \begin{tabular}{lllll}
 \toprule
 \emph{IV} & \emph{Test case content} & \emph{Term extraction} & 
 \emph{IR Model} \\
 \midrule
 & Full test case & Tokenization & VSM \\
 \emph{Values} & Test case except test steps & Stop words & LSA \\
 & Acceptance criteria and objective & Stemming \\
 \bottomrule
 \end{tabular}
\end{table}

As described in Section~\ref{sub:object_description}, test cases consist of 
categorization and execution information, containing information that is not 
necessarily connected to the to-be-tested feature. For example, assumptions, 
initialization information and test steps may refer to pre-requisite 
functionality that is verified by another dedicated test case. Hence, it is 
necessary to identify the test case content that is the most effective, i.e. 
encode the most information on a feature with the least amount of noise.

Term extraction refers to pre-processing techniques applied incrementally on 
the analyzed texts, where the simplest form, tokenization and stop-word removal, 
can be extended by stemming~\citep{falessi_empirical_2013}. Furthermore, 
techniques exist that aim specifically to improve IR applied on source code, 
e.g. by splitting~\citep{enslen_mining_2009,dit_can_2011} 
or expanding identifiers~\citep{hill_amap:_2008,lawrie_expanding_2011} or a 
combination of both~\citep{guerrouj_tidier:_2011,corazza_linsen:_2012}. Even 
though these techniques seem to improve the performance of IR, they have been 
implemented for and validated against only a subset (Java, C, C++) of the 
programming languages we encountered in a product (see 
Section~\ref{sub:setup1step3}). Therefore, we decided against adding them as a 
manipulated variable in our experiment.  

An IR model evaluates the semantic similarity between text 
documents~\citep{falessi_empirical_2013}. The vector space model 
(VSM)~\citep{salton_vector_1975} and latent semantic analysis
(LSA)~\citep{deerwester_indexing_1990} both represent documents as 
term-frequency vectors and consider the distance between these vectors as 
semantic dissimilarity. LSA, an extension to the VSM, considers co-occurrence 
of terms~\citep{falessi_empirical_2013} in order to address synonymy (the same 
meaning - different terms) and polysemy (different meanings - the same 
term)~\citep{deerwester_indexing_1990}. Note that LSA is a parameterized model, 
increasing the number of independent variables.

% \subsection{Dependent variables}\label{sub:depvar}
% [So what are our dep vars here... similarity metrics or precision/recall or 
% rankings? Differs from setup to setup. Hence write setups first, then 
% generalize here. Also consider adding run-time as a dep var.]

%Similarity metric (cosine) is dv, but during the different setups, different 
%approaches on how to evaluate the results were implemented. 

%Concretely, if we have two feature vectors $A_i$ and $A_i'$ 
%hat differ by the activation of $F_i$, then we evaluate the difference 
%between $Sim(V(A_i), T(F_i))$ and $Sim(V(A_i'), T(F_i))$.

\subsection{Validity threats}\label{sub:validitythreats}
We use \cite{wohlin_experimentation_2000} to structure this analysis and 
discuss the threat categories in the suggested priority for empirical software 
engineering experiments: internal, external, construct and conclusion threats. 
For each category, we discuss only those threats that apply to the experiment 
design formulated in this section.

%- while link from feature to source code is established by an objective 
%mechanism, link feature to test case is established by a test engineer.
%- only a subset of potential indep variables are considered, see 
%Falessi/Biggers who did a more extensive selection

\subsubsection{Internal validity} 
The \emph{instrumentation} threat refers to the effects caused by the artifacts 
used for experiment implementation. The impact of the used artifacts is a 
central aspect of the study overall, but in particular of the experimental 
design and its implementation in three setups. As such, we are explicitly 
addressing and discussing instrumentation and its effect in each setup.

The \emph{selection} threat refers to the variation of the chosen objects in 
the experiment. The feature chains are determined by the possibility to 
establish an association between a particular feature and the test cases that 
verify that feature. As such, the selection of features is biased towards those 
for which the tasked test engineers could create such a mapping. We have 
however no indication that the non-random selection of features introduced a 
systematic bias towards product variants and their artifacts or types of system 
test cases.

\subsubsection{External validity}
The \emph{interaction of selection/setting and treatment} threat refers to the 
selection of objects from a population that is not representative for the 
population to which we want to generalize. We address this threat by sampling 
objects from an industry setting which is the population to which we generalize 
our results. 

\subsubsection{Construct validity}
\emph{Inadequate pre-operational explication of constructs} refers to the 
threat of insufficiently defining the theoretical constructs of the experiment. 
We address this threat by formalizing the relationships of the objects, i.e. 
the feature chain and its components, in the experimental design and using this 
formalization in the execution of the three experiments. However, there is a 
threat that the mapping of test cases to features, necessary for constructing a 
feature chain, was biased by the understanding and experience of a single 
tester. We addressed this by involving two test engineers in this process.

A \emph{mono-operation bias} exists in an experiment if the limited number of 
independent variables, objects or treatments leads to an incomplete picture of 
the theory. While we aimed at selecting a sensible variation of independent 
variables, the choice was eventually limited by the practical constraints of 
implementing the experiment on industry grade data sets, that addressed 
external validity threats. This is a trade-off between conflicting validity 
threats one has to consider when performing applied 
research~\cite{wohlin_experimentation_2000}. Other studies focus on a broader 
coverage of IR configurations, e.g. ~\cite{falessi_empirical_2013} and 
\cite{biggers_configuring_2014}.

\subsubsection{Conclusion validity}
There is a moderate threat of \emph{low statistical power} w.r.t. the 
conclusions made in Experimental Setup 3. The cost for establishing an 
observation (feature chain) considerably limited the number of observations 
that could be made with reasonable effort, also from the industry participants. 

We address the threat of \emph{violated assumptions of statistical tests} by an 
in-depth analysis of the experimental design and choice of statistical 
techniques (see Section~\ref{sub:setup3lessons}). 

While the implementation of corpus creation, term extraction, similarity 
analysis and evaluation differed in the three experimental setups (see 
Figure~\ref{fig:experimental_setups}) due do their varying objectives, we 
consider the threat of \emph{reliability of treatment implementation} low. The 
implementation of the four steps did not change within a experimental setup and 
is available in the supplementary 
material~\citep{unterkalmsteiner_supplementary_2014} for reference.

\section{Experimental setups and evaluation}\label{sec:expsetup}
\begin{figure}
 \begin{center}
  \includegraphics[scale=0.6]{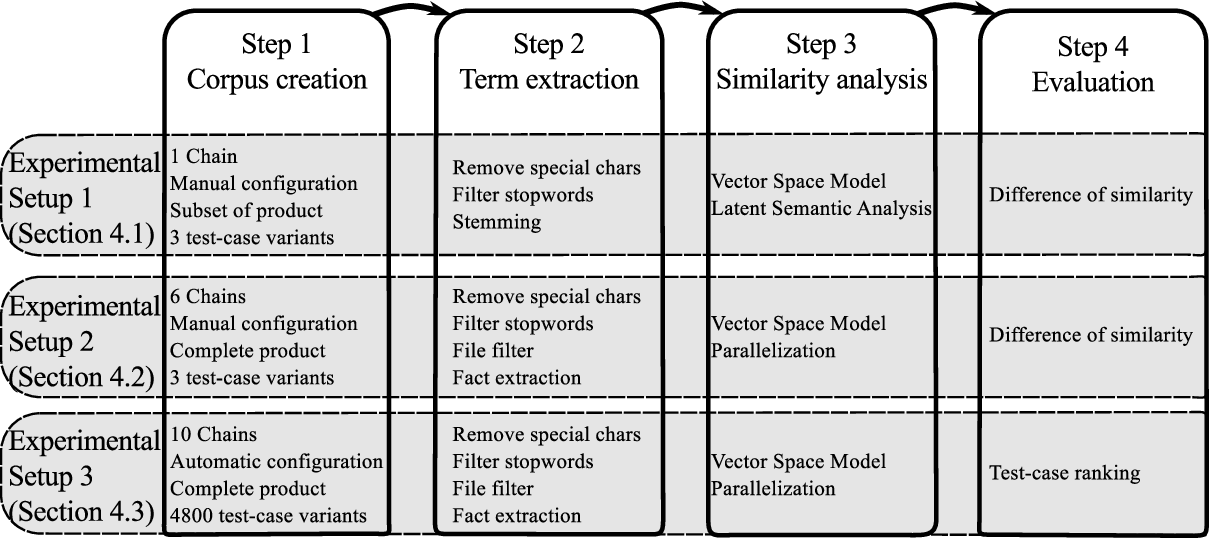}
 \end{center}
\caption{General experiment procedure (horizontal axis) and changes between the 
three experimental setups (vertical axis)}
\label{fig:experimental_setups} 
\end{figure}
In this section we illustrate three major experimental setups whose 
implementation is guided by the design presented in 
Section~\ref{sec:problem_definition}. Each setup had a particular objective and 
its implementation lead to lessons learned, motivating and shaping the 
objectives of the subsequent setup.

% To enable evaluation we need to define a ground truth. We can exploit the 
% fact that features can be deactivated from the configuration file, effectively 
% removing them from the product. By tracing the impact of a particular feature 
% activation switch to the affected product artifacts we know the feature 
% location. Then, the feature is mapped by test engineers to a set of 
% test cases which they deem to be relevant. By this procedure we establish a 
% ``chain'' (see Figure~\ref{fig:chain}), from feature realization in the source 
% code, over the feature activation switch to the set of relevant test cases.

The execution of the experiment follows the general flow shown in 
Figure~\ref{fig:experimental_setups} in all three setups. In Step 1, we create 
the corpus for a chain, consisting of the configured product artifacts and 
test-cases. In Step 2, the corpus is textually pre-processed, which removes 
noise and prepares the data for the subsequent analysis in Step 3. We perform 
the evaluation of the IR technique in Step 4. With each experimental setup, 
the steps are refined to address issues and lessons learned from the 
predecessor. An overview of the differences between the three experimental 
setups is shown in Figure~\ref{fig:experimental_setups}, while the details are 
discussed in the respective subsections.

\subsection{Experimental Setup 1: Pilot Experiment}\label{sub:setup1}
The motivation for preparing a pilot experiment is to get an intuition whether 
IR techniques can be used to solve the test case selection problem. Therefore, 
we designed an experimental setup that tests the sensitivity of IR. In other 
words, we want to determine whether the signal provided by the test cases is 
strong enough to differentiate between a feature-activated and a 
feature-deactivated corpus. 

Formally, using the notation introduced in 
Section~\ref{sub:object_description}, we have in Experimental Setup 1 one 
feature chain $I_1$ that includes feature $F_1$ and test case $T_1$. 
Therefore, we have two feature vectors $A_a=[1, a_2, a_3,\ldots, a_{fn}]$ and 
$A_d=[0, a_2, a_3,\ldots, a_{fn}]$, resulting in two system variants 
$V(A_a)=\{C_1,\ldots,C_{cn}\}$ and $V(A_d)=\{C_1,\ldots,C_{cm}\}$. We evaluate 
whether $sim(T_1, V(A_a)) > sim(T_1, V(A_d))$, where $T_1$ is the 
test case mapped to feature $F_1$. As for independent variables, we manipulate 
the content of $T_1$ and the IR model $sim$ (see Table~\ref{tab:iv}).
Figure~\ref{fig:experimental_setups} illustrates the configuration of each step 
in this setup, further detailed in 
subsections~\ref{sub:setup1step1}~-~\ref{sub:setup1step4} and in
Figure~\ref{fig:ir}. 

% The overall idea of the procedure is to create two 
% product configurations, one with the feature activated, the other with the 
% feature deactivated, creating two distinct corpora (Step 1). After 
% pre-processing the corpora (Step 2), we use the test case as a query to locate 
% the feature in the two corpora (Step 3), using VSM and LSA respectively. We 
% intend to measure a difference in location performance between the feature 
% activated and deactivated corpus (Step 4). Such a measurable difference would 
% indicate that IR techniques can be used for test case selection. 

The overall objectives of the pilot experiment are to:
\begin{itemize}
 \item increase the understanding of the analyzed data (product artifacts, 
test cases)
 \item identify and evaluate IR tools which can be used to implement and 
execute the experiments
 \item provide evidence on whether this approach is generally 
feasible or not
\end{itemize}

We discuss in Sections~\ref{sub:setup1step4}~and~\ref{sub:setup1lessons} to 
what extent these objectives have been achieved.

\subsubsection{Step 1 - Corpus creation}\label{sub:setup1step1}
The product is configured by a collection of GNU 
autotools~\citep{calcote_autotools:_2010} scripts and built with GNU 
make~\citep{feldman_make_1979}. In a typical product build, several hundred 
software packages are fetched from the version control system server, 
configured, compiled and packaged into an image that can be installed onto the 
respective camera model.

For the pilot experiment, we selected one feature that is implemented in a 
relatively small software package (368 files). On activation, the feature 
affects 10 source code files (8 modifications and 2 additions). The build system 
uses filepp\footnote{\url{http://www-users.york.ac.uk/~dm26/filepp}}, a file 
pre-processor, to apply configuration switches on plain text and HTML files. By 
manually invoking filepp on the software package that implements the user 
interface, we generated two variants with 10 distinct and 358 shared files. 

% As described in Section~\ref{sub:object_description}, a test-case contains 
% categorization and execution information. We created three variants of 
% test-cases:
% \begin{itemize}
%  \item the complete test-case 
%  \item the test-case without the test steps 
%  \item only the test objective and acceptance criteria
% \end{itemize}

\subsubsection{Step 2 - Term extraction}
We applied the following pre-processing steps on both the source code 
and test case variants:
\begin{itemize}
 \item filtered numeric, punctuation and special characters\footnote{using GNU 
sed, \url{http://www.gnu.org/software/sed/}}
 \item filtered stop-words; English, and HTML/javascript keywords
 \item performed word stemming
\end{itemize}

Stop-word filtering and word stemming were implemented using the generic data 
mining tool Rapidminer\footnote{\url{http://www.rapidminer.com}, version 5}, 
which was also used in the next step.

\subsubsection{Step 3 - Similarity analysis}\label{sub:setup1step3}
\begin{figure}[t]
 \begin{center}
  \includegraphics[scale=0.8]{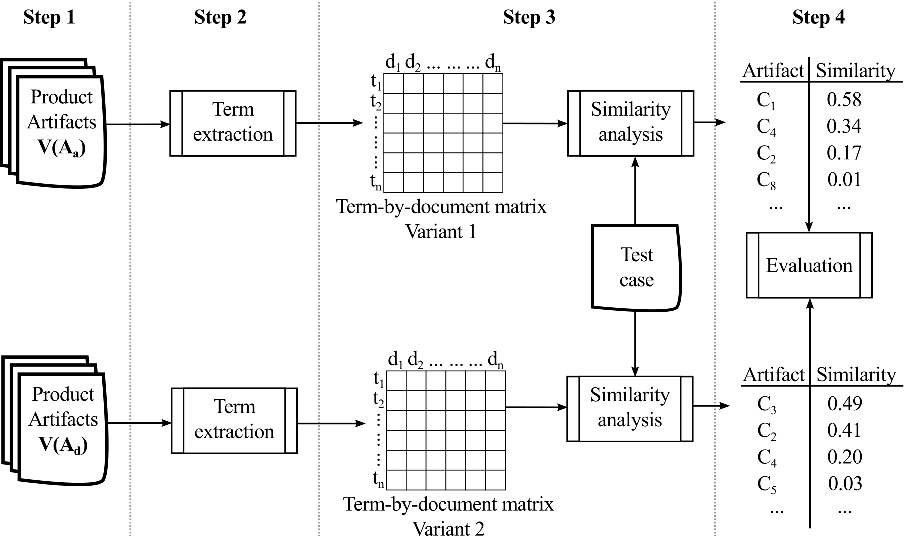}
 \end{center}
\caption{Pilot setup details}
\label{fig:ir} 
\end{figure}

Figure~\ref{fig:ir} illustrates the pilot setup in more detail, emphasizing the 
fact that we are working with two corpora ($V(A_a)$ and $V(A_d)$ with 
activated and deactivated feature).

When creating the term-by-document matrix in Step 3, one has to decide on the 
document granularity, i.e. on how the corpus is divided into documents. 
Commonly, when analyzing similarity in source code, documents are defined as a
method~\citep{marcus_information_2004,cleary_empirical_2009}, a 
class~\citep{antoniol_recovering_2002} or a 
file~\citep{marcus_recovering_2003}. We defined a file as a document due to the 
heterogeneity of the source code (markup, procedural, object oriented 
languages).

\begin{figure}
 \begin{center}
  \includegraphics[scale=0.8]{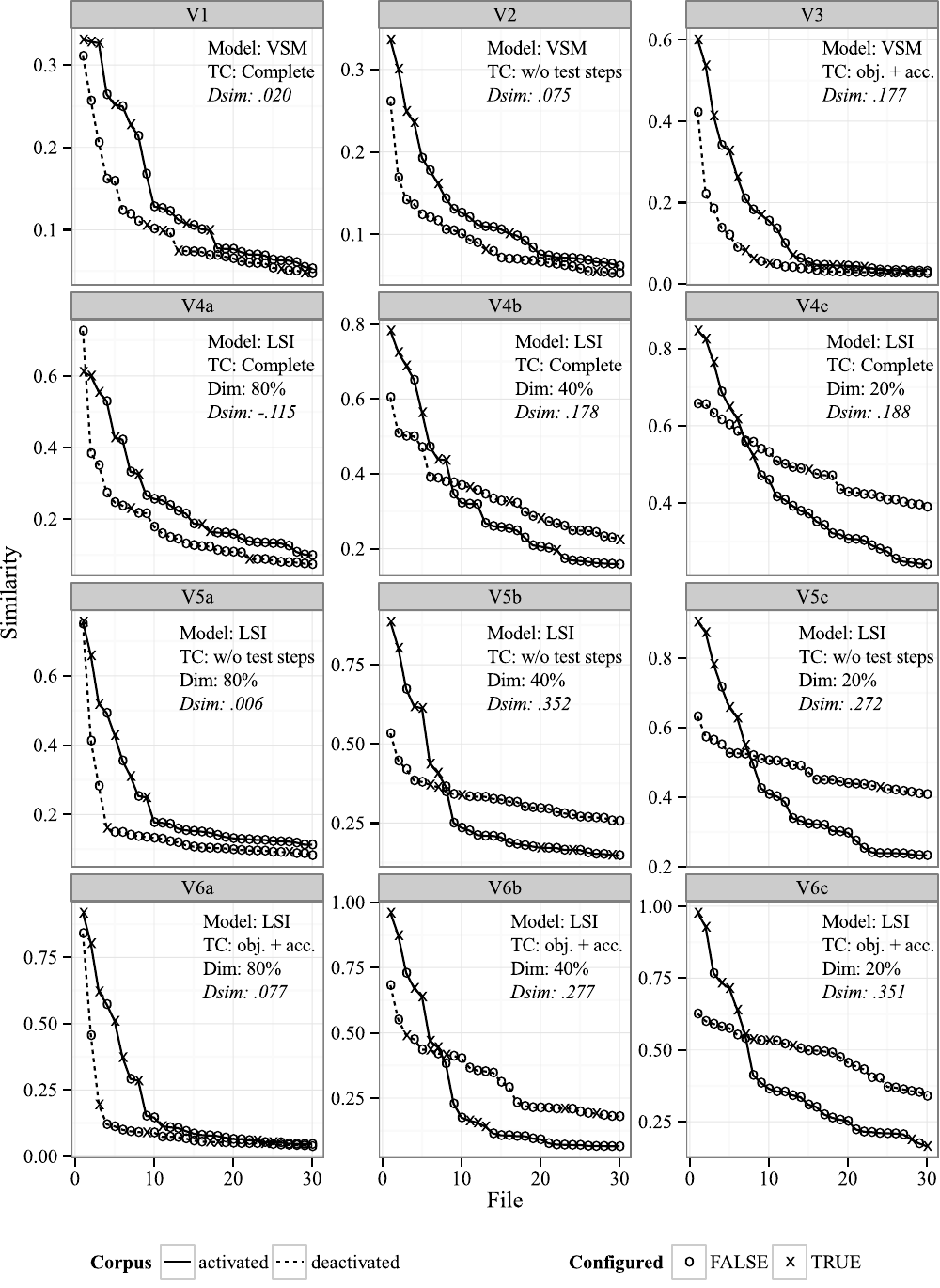}
 \end{center}
\caption{Pilot experiment results}
\label{fig:gatekeeper_pilot} 
\end{figure}

We applied VSM and LSA on the corpora, resulting in a list of ranked 
documents according to their similarity to the test case. 
Figure~\ref{fig:gatekeeper_pilot} shows the results of 12 runs of the pilot 
experiment: with the VSM, we varied the test case content (V1-V3), whereas with 
LSA, we also varied dimension reduction (V4abc-V6abc). For each run, we plotted 
the cosine similarity of the first 30 ranked documents (for all runs, the 
similarity values approach values below 0.1 after 30 documents and showing all 
documents would render the plots more difficult to read). Furthermore, 
documents that differ between the two corpora, i.e. that are affected by 
activating/deactivating the feature, are marked with an ``x''-symbol.

Looking at Figure~\ref{fig:gatekeeper_pilot}, it can be observed that for all 
runs, in the feature activated corpus (solid line), configured documents 
are ranked first (e.g. 4 in V1, 2 in V4a, etc). This shows that, for this 
particular corpus and test case, all IR techniques and combinations are able to 
locate the feature in question. However, we are interested in 
differences between corpora (solid and dashed lines). These differences are 
visible in runs V1-V3 (VSM), i.e. the maximum similarity of the feature 
activated corpus is higher than the one of the feature deactivated corpus. 
Furthermore, the solid line (activated corpus) is consistently above the dashed 
line (deactivated corpus). This indicates that, at least visually, we can 
differentiate between the two corpora. This changes in runs V4-V6 
(LSA), rendering the visual differentiation more difficult. 
We can observe that the maximum similarity in run V4a is higher in the 
feature deactivated corpus higher than in the feature activated corpus. 
Furthermore, in runs V(456)b and V(456)c the solid line intersects the dashed 
line at around file 8, indicating that the overall calculated similarity 
between test case and feature deactivated corpus is higher than for the feature 
activated corpus.

These results are promising as they indicate that a differentiation between a 
feature activated and deactivated corpus is possible, i.e. that there is a 
discerning signal in this particular test case and IR techniques are 
able to detect that signal. However, in order to determine which combination of 
test-case component and IR technique performs best, we need to define a measure 
that is able to express the visually determinable difference as a numerical 
value, as shown next.

\subsubsection{Step 4 - Evaluation}\label{sub:setup1step4}
In Section~\ref{sub:setup1} we defined an experimental setup with two corpora 
in order to test whether IR techniques can differentiate, given a test case, 
between the two product variants. This is motivated by our aim for test case 
selection, which needs to determine the existence/non-existence of a feature, 
but not necessarily its exact location.

However, this also means that the traditional class of performance measures, 
e.g. precision and recall, based on a confusion matrix, cannot be used for the 
following reason. In a binary classification problem, the confusion matrix 
contains four classes: true positives, true negatives, false positives, and 
false negatives. Given an arbitrary threshold value for the calculated 
similarity between document (source code) and query (test case), one can define
a confusion matrix for the corpus where the feature is activated. However, 
given a corpus where the feature is deactivated, the number of true 
positives and false negatives, i.e. the number of relevant documents, is by 
definition zero. In other words, no classification takes place in the feature 
deactivated corpus. Hence, measures based on the confusion matrix are 
unsuitable in this case.

Therefore, we develop a measure based on similarity and the given ground 
truth. 
Similarity, a direct measure as opposed to the derived precision/recall pair, 
has been used by~\cite{falessi_empirical_2013} to construct a credibility 
measure. Following this example, we construct a differential similarity measure,
\begin{equation}\label{dsim}
 Dsim=max(sim_{activated | configured}) - max(sim_{deactivated})
\end{equation}
which is the difference of the maximum similarity values of the two corpora. 
Note that the maximum similarity for the feature activated corpus is taken from 
the artifact that was actually affected by the feature activation, i.e. is 
configured.
Since the calculated similarity between two documents ranges from 0 to 1, 
the range for Dsim is between -1 and 1. 1 indicates a perfect 
true differentiation, 0 no differentiation at all, and -1 a perfect false 
differentiation. A false differentiation occurs when the IR technique fails to 
connect the test case to the correct, i.e. configured, source code artifacts.

It would be possible to construct sophisticated measures, incorporating more 
of the distinguishing features of the curves in 
Figure~\ref{fig:gatekeeper_pilot} and the characteristics of feature locations. 
For example, one can assume that features are implemented in a small number of 
files out of a larger set. Then, the similarity for this small number of files 
would be large, followed by a drop-down and a quick convergence to 0 (as 
it can be observed in all curves of the feature activated corpus in 
Figure~\ref{fig:gatekeeper_pilot}). For the feature deactivated corpus, one 
would expect only a small drop-down in similarity and a rather slow convergence 
to 0, as it can be observed in V(456)c in Figure~\ref{fig:gatekeeper_pilot}. 
These features of the curves, drop-down and convergence of similarity 
measures, could be used to define more accurate differential similarity 
measures. However, this approach would be threatened by an over-fitting of the 
measure definition to the given data, resulting in a measure that would 
represent well the current, but not the future data. Hence, we choose the simple 
$Dsim$ measure that relies only on the maximum similarity values.

\subsubsection{Discussion and Lessons Learned}\label{sub:setup1lessons}
We start this section by discussing the results of the evaluation and then 
elaborate on the lessons learned from the implementation of Experimental Setup 
1, addressing RQ-1.

Looking at Figure~\ref{fig:gatekeeper_pilot}, run V5b, V6c and V6b achieve the 
largest $Dsim$ value, indicating that LSA performs better than VSM (V1-V3). In 
general, we can observe that the more specific the test case content, the 
larger the $Dsim$ value, i.e. using the complete test case is consistently 
worse (for both VSM and LSA) than using a subset of the test case. This is an 
important result as it indicates that test cases contain noise that can be 
reduced by a selective removal of test-case content.

In this pilot setup, we test our idea of using IR techniques for test case 
selection. The results, illustrated in Figure~\ref{fig:gatekeeper_pilot}, 
indicate that for this particular chain and subset of documents, we can indeed 
differentiate between a feature activated and a feature deactivated variant of 
a product, using the corresponding test case as a probe. However, further 
experimentation as described in Section~\ref{sub:setup2} is required to 
investigate the scalability and generalizability of the approach.

Looking at the number of published studies in the field of feature 
location~\citep{dit_feature_2011}, or in other areas where IR techniques are 
applied, e.g. traceability research~\citep{borg_recovering_2013}, one would 
assume that literature provides both technical and methodological guidance to 
implement and evaluate IR-based solutions, adapted for the particular problem 
context. 
However, very few studies provide enough detail to implement their 
proposed approach. None of the 14 publications that study textual feature 
location techniques based on VSM and/or LSA reviewed by~\cite{dit_feature_2011} 
provides prototype implementations. Few mention existing tools that were used 
to implement parts of the proposed approaches. 
~\cite{abadi_traceability_2008} and \cite{gay_use_2009} use Apache 
Lucene~\citep{the_apache_software_foundation_apache_2014} as the VSM 
implementation, and ~\cite{cleary_empirical_2009} mention 
SVDPACKC~\citep{berry_svdpackc_2014} as an implementation for singular value 
decomposition (SVD), a procedure required for LSA. \cite{lormans_can_2006} use 
the Text to Matrix Generator toolbox~\citep{zeimpekis_tmg:_2006}, a Matlab 
module for text mining. 

However, none of these tools lends itself to set up an experimental environment
that allows one to explore data and technologies. Therefore, we 
implemented the process in Figure~\ref{fig:experimental_setups} with the 
off-the-shelf data-mining tool Rapidminer. The tool provides text processing 
capabilities and the necessary operators to implement VSM and LSA models. 
Furthermore, analyses can be constructed through the component-based graphical 
process designer, allowing for quick exploration of ideas. However, this 
convenience limits flexibility, e.g. by providing only a limited number of term 
weight functions\footnote{Nevertheless, standard functionality can be extended 
through Rapidminers' plugin mechanism}. Furthermore, the execution time for a 
single experiment run in the pilot does not scale to the amount of data in an 
actual chain. We implemented therefore the pilot process and the 
following experimental setups with shell scripts the statistical software 
package R~\citep{crawley_r_2007}, available in the supplementary 
material~\citep{unterkalmsteiner_supplementary_2014}. While there exist efforts 
to reduce the barrier for conducting IR experiments, e.g. with 
TraceLab~\citep{cleland-huang_grand_2011,dit_supporting_2014}, researchers 
require also the flexibility of general purpose statistical software. For 
example, \cite{dit_configuring_2013} piloted their approach with TraceLab, 
implemented the final experiment~\citep{panichella_how_2013} however with R. 
This mirrors our experience with Rapidminer and R, where the former provides 
the ease of use for quick experimental exploration while the latter enables 
flexible configuration and batch processing.

% see http://radimrehurek.com/gensim/wiki.html#latent-semantic-analysis
%for LSI implementation in python that works also on multicore/distributed 
%systems
% see http://radimrehurek.com/phd_rehurek.pdf for a discussion on scalability 
% of NLP, in particular LSI and LDA

The studies reviewed by \cite{dit_feature_2011} and \cite{borg_recovering_2013} 
usually describe process steps, such as corpus creation, indexing, query 
formulation, and ranking. However, this information alone is insufficient to 
re-implement the proposed approaches as the IR techniques' behavior is defined 
by a set of parameters~\citep{thomas_impact_2013}, including pre-processing 
steps, input data, term weights, similarity metrics and algorithm-specific 
variations such as the dimensionality reduction factor $k$ for LSA. Choosing 
$k$ is a matter of experimenting with the data at hand to identify the value 
that provides the best retrieval performance~\citep{deerwester_indexing_1990}. 
%Note to myself: page 396 in their paper
When comparing different algorithms, it is necessary to identify the optimal 
parameters in order to provide a ``fair'' 
comparison~\citep{hooker_testing_1995}. 
This process is referred to as parameter tuning (see for example 
\cite{lavesson_quantifying_2006} and \cite{arcuri_parameter_2011}). 
Few of the studies reviewed by \cite{dit_feature_2011} and 
\cite{borg_recovering_2013} tune $k$ to their LSA application. Some pick $k$ 
based on experiences from previous studies (e.g. from 
\cite{deerwester_indexing_1990} and \cite{dumais_lsi_1992}), which is however 
a questionable practice: these early studies use benchmark data-sets 
unrelated to the software engineering context. Furthermore, SVD on large, 
sparse matrices is a computationally expensive operation. Iterative 
implementations (e.g. \cite{berry_svdpackc_2014}) using the Lanczos 
algorithm~\citep{cullum_lanczos_2002} approximate the singular values, 
effectively reducing the runtime of LSA by limiting the number of dimensions. 
Computational cost should however not be a hidden driver for parameter 
selection as it biases the comparison~\citep{hooker_testing_1995}.

Few studies motivate, empirically or theoretically, their choice of 
$k$. A notable exception is \cite{de_lucia_recovering_2007} where the 
performance of LSA is systematically evaluated. Since the document space in 
their experiment was small (150), it was computationally not costly to vary $15 
\leq k \leq 150$. The authors observed that LSA performance, measured with 
precision and recall, did not vary much when $k$ was set between 30\%-100\% of 
the document space, attributing this behavior to the small number of documents 
used in their experiments, and leading to the conclusion that $k$ should be a 
user-configurable parameter in their proposed tool. 
\cite{poshyvanyk_combining_2006} worked with a larger document space (68,190), 
varied $k$ however only between 0.4\%-2.2\% ($k=[300, 500, 750, 1500]$), 
motivating their decision that larger values for $k$ are impractical to 
compute. They also observed that varying $k$ did not significantly 
influence LSA performance. 

However, concluding from these results that an optimal value for $k$ should be 
between 30-500 would be wrong. \cite{de_lucia_recovering_2007} observed a
decline in performance when $k$ was set to less than 10\% of the document 
space. \cite{poshyvanyk_combining_2006} never reached that percentage due to 
the computational cost involved in SVD. Not optimizing model parameters can 
lead to contradicting conclusions on which IR technique performs better, as the 
following example illustrates. ~\cite{oliveto_equivalence_2010} compare, among 
other models, LSA with Latent Dirichlet Allocation (LDA). While $k$ for LSA is 
not reported, the number of topics in LDA is varied between 50 and 300. The 
authors conclude that ``the LDA-based traceability recovery technique
provided lower accuracy as compared to other IR-based 
techniques''~\citep{oliveto_equivalence_2010}. ~\cite{asuncion_software_2010}, 
using the same standardized data set (EasyClinic\footnote{
http://web.soccerlab.polymtl.ca/tefse09/Challenge.htm}), parameterize LSA 
with $k=10$ and LDA with $t=10$, and conclude that LDA performs better 
than LSA, i.e. the opposite of~\cite{oliveto_equivalence_2010}. These 
contradicting results, also observed by \cite{grant_using_2013}, illustrate the 
importance of parameter optimization.
%NOTE 20140916: In "Using heuristics to estimate an appropriate number of latent
%topics in source code analysis (2013)", the same example with the Asuncion and 
%Oliveto study is made!

% Examples of NLP studies with / without detailed context and tool 
% description/availability

%falessi_empirical_2013: Section 4.2.3, ANTARCTICA tool not available (only 
%link to presentation), no implementation details. Number of requirements 
%given, but not size, i.e. amount of processed and analyzed data --> important 
%to understand runtime.

%``Semantic clustering: Identifying topics in source code'': Small Systems 
%(Figure 6). Steps of method explained, tool (Hapax) mentioned, but no 
%implementation details or link to implementation given.

%% NUMBER OF DIMENSIONS in LSI::
% ``Semantic clustering: Identifying topics in source code'': Section 2
% ``Combining Formal Concept Analysis with Information Retrieval for Concept 
%Location in Source Code'': Section 5.1.2
% ``On the Influence of Latent Semantic Analysis Parameterization for Bug 
%Localization'': Section 2.4

\subsection{Experimental Setup 2: Scaling up}\label{sub:setup2}
While Experimental Setup 1 focused on studying the feasibility of using IR 
techniques for test case selection, this setup aims at evaluating the 
scalability of the approach. Concretely, the objectives are to study its 
behavior with respect to:
\begin{itemize}
 \item computational scalability of the 4 steps (corpus creation, term 
extraction, similarity analysis and evaluation), and
 \item accuracy, i.e. a larger corpus means that the feature location task 
becomes more difficult since there are many more irrelevant than relevant
documents 
\end{itemize}

In this setup, we apply the piloted approach shown in Section~\ref{sub:setup1} 
on a set of corpora created from the complete product. Formally, we have in 
Experimental Setup 2 six feature chains $I_{1\ldots6}$ that include six 
features $F_{1\ldots6}$ and six test cases $F_{1\ldots6}$. Therefore, we have 
12 feature vectors $A_{a1\ldots6}$ and $A_{d1\ldots6}$ resulting in 12 system 
variants $V(A_{a1\ldots6})$ and $V(A_{d1\ldots6})$. We evaluate whether 
$sim(T_{1\ldots6},V(A_{a1\ldots6})) > sim(T_{1\ldots6},V(A_{d1\ldots6}))$. 
As for independent variables, we manipulate the content of $T_{1\ldots6}$. The 
characteristics of this setup are summarized in 
Figure~\ref{fig:experimental_setups} and detailed in 
subsections~\ref{sub:setup2step1}~-~\ref{sub:setup2step4}. We 
report on the lessons learned in Section~\ref{sub:setup2lessons}.

\subsubsection{Step 1 - Corpus creation}\label{sub:setup2step1}
We selected six features, resulting in chains consisting of between 67,238 and 
67,334 files, depending on the particular feature activation state. As in 
Experimental Setup 1, we created the different product configurations manually. 
However, in contrast to the limited set of files used in Experimental Setup 1, 
the complete product contains source code written in various programming 
languages, and build, configuration and documentation files. Furthermore, we had 
to consider more configuration mechanisms. In Experimental Setup 1, the feature 
activation/deactivation could be performed with filepp alone since only a 
subset of the product (the one configured with filepp) was considered for the 
corpus. However in this setup, GNU autotools mechanisms (i.e. depending on 
configuration, adding or removing files to the build process) and 
pre-processor mechanisms from the GNU compiler tool-chain needed to be 
considered in addition. In this setup we did not build the product (this will 
be explored in Experimental Setup 3 in Section~\ref{sub:setup3}), but traced 
the impact of feature activation/deactivation and implemented it manually in 
the product artifacts. 

For each feature chain, we randomly selected one test-case from the pool of 
test cases identified by the test engineers as relevant for the particular 
feature. As in the pilot experiment, we created three test case variants that 
differed by the amount of included information. 

\subsubsection{Step 2 - Term extraction}\label{sub:setup2step2}
With the increase in size but also the variety of the corpus, term extraction 
became more challenging. We decided to eliminate term stemming as the 
processing time for one feature chain amounted to 20 hours. This likely 
increased the number of terms in the term-by-document matrix. On the other 
hand, we added a filter that removed binary files from the corpus and added a 
fact extraction process, inspired by~\cite{poshyvanyk_combining_2007}. We used 
srcML~\citep{maletic_source_2002} to extract 
identifier names, comments and literals from C and C++ source code, reducing 
the amount of irrelevant terms in the corpus. Concretely, for the kernels' 
source code files, the term count could be reduced from 237,393 to 
195,019 (a reduction of 18\%) with fact extraction.

\subsubsection{Step 3 - Similarity analysis}\label{sub:setup2step3}
To implement the term-by-document matrix we used an R text-mining 
package~\citep{feinerer_text_2008}, which provides efficient means to represent 
highly sparse matrices. This is required to be able to efficiently process a 
large corpus in memory. Furthermore, we improved the computational performance 
in two ways.

First, we optimized the calculation of similarity values by changing our 
initial, iterative, implementation into a vectorized form, exploiting the 
performance advantages of array programming~\citep{iverson_notation_1980}.
This reduced the similarity calculation using the VSM for one chain (corpus with 
60,000 files) from 20 hours to 8 hours. 

Second, we changed our implementation of VSM to support parallelization, 
allowing us to distribute the workload among several CPUs. We chose to use the 
snowfall package~\citep{knaus_snowfall:_2013} for this task as it allows to 
choose at run-time, without changing code, whether to run in cluster or in 
sequential mode, which is useful for testing the implementation. With the use 
of a cluster consisting of 8 cores, we could reduce the computation 
time for one chain from 8 to approximately 1 hour.

\subsubsection{Step 4 - Evaluation}\label{sub:setup2step4}
We used the Dsim measure, introduced in Section~\ref{sub:setup1step4}, to 
evaluate the performance of VSM to differentiate between the activated and 
deactivated corpus using a test-case. Table~\ref{tab:setup2dsim} shows the 
results of the 18 experiment runs: 6 chains and 3 test-case variants each. 
Recall that for a true differentiation, the Dsim measure ranges between 0 and 
1. 

\begin{figure}
 \centering
 \subfloat[Chain 2]{\label{subfig:chain4}
   \includegraphics[scale=0.7]{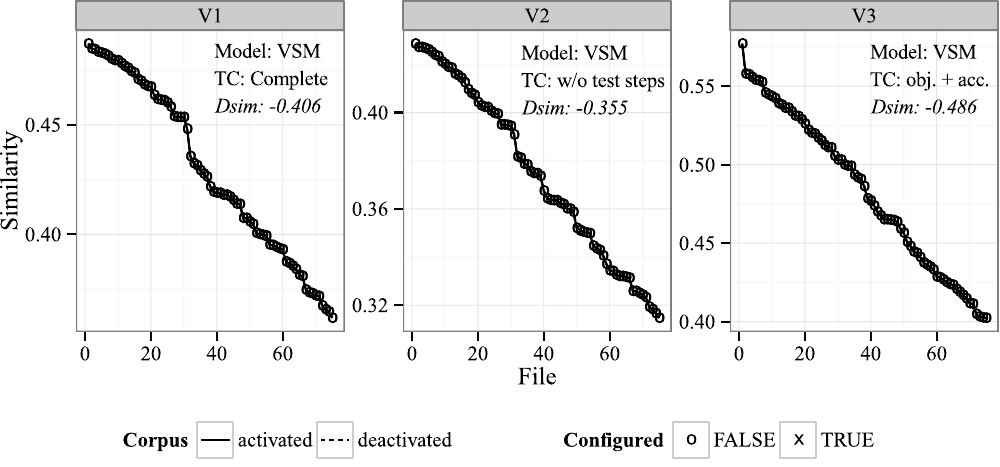}}
 \\
 \subfloat[Chain 3]{\label{subfig:chain9}
  \includegraphics[scale=0.7]{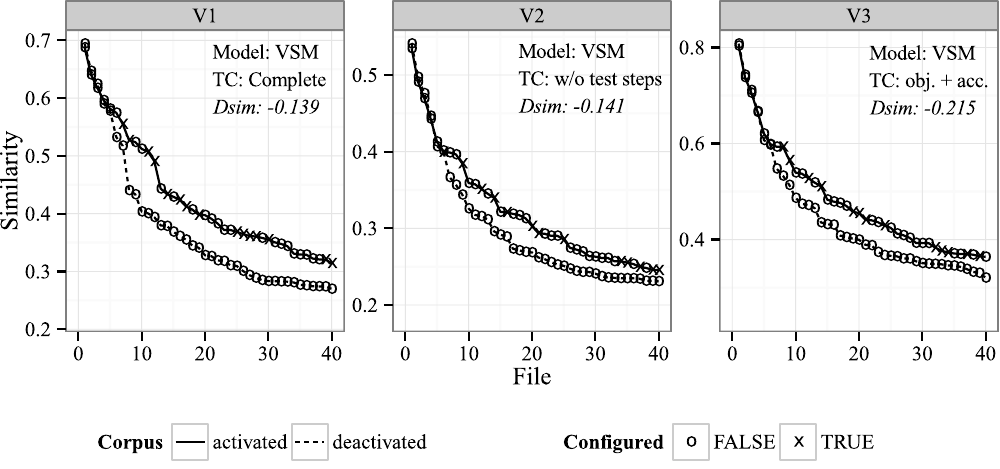}}
  \\
  \subfloat[Chain 6]{\label{subfig:chain10}
   \includegraphics[scale=0.7]{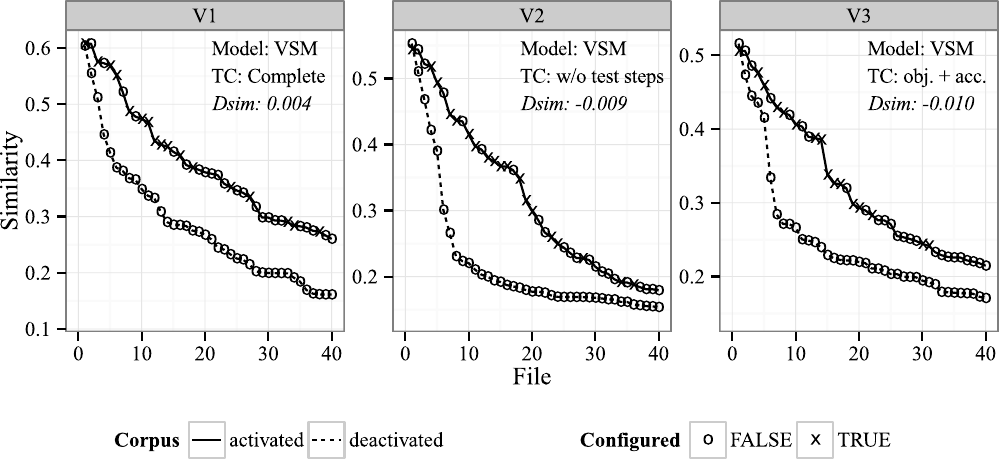}}
\caption{Results on three complete product variants}
\label{fig:setup2results} 
\end{figure}

Looking at Table~\ref{tab:setup2dsim}, the results from chain 1 indicate that a 
true differentiation is possible. This confirms the results from Experimental 
Setup 1, which consisted of the same feature and test-case, however with a 
subset of the product artifacts. On the other hand, the Dsim measure of chains 
2-6 indicates that the test-cases cannot differentiate between the feature 
activated and deactivated corpus. Looking at chain 2, shown in 
Figure~\ref{subfig:chain4}, we observe that there 
is no dissimilarity between the feature activated and deactivated 
corpus among the first 70 files. For chains 4 and 5 the curves look similar and 
are therefore not shown in Figure~\ref{fig:setup2results}. Common to these 
three chains is the low number of files that are affected by a feature 
activation (between 2 and 7) and the change is minimal (typically a parameter 
replacement in a configuration file). In these cases, IR techniques are 
unlikely to work for test case selection as the changes between variants are 
minimal. 

The situation is different for chains 3 and 6. Even though the Dsim measures 
in Table~\ref{tab:setup2dsim} indicate no differentiation, 
Figure~\ref{subfig:chain9} indicates a slight, and Figure~\ref{subfig:chain10} 
a more pronounced difference between activated and deactivated corpus. 
This suggests that the Dsim measure discards too much information by taking 
only the maximum similarity values into consideration.

% Mapping of chain id's to id's in this paper:
% 1 -> 1
% 2 -> 4
% 3 -> 9
% 4 -> 3
% 5 -> 5
% 6 -> 10
\begin{table}
 \caption{VSM results using the Dsim measure}\label{tab:setup2dsim}
 \centering
 \begin{tabular}{lcccccc}
 \toprule
 \emph{Variant / Chain} & 1 & 2 & 3 & 4 & 5 & 6 \\
 \midrule
 \emph{V1} & 0.120 & -0.406 & -0.139 & -0.490 & -0.259 & 0.004 \\
 \emph{V2} & 0.170 & -0.355 & -0.141 & -0.504 & -0.244 & -0.009 \\
 \emph{V3} & 0.228 & -0.486 & -0.215 & -0.517 & -0.205 & -0.010 \\
 \bottomrule
 \end{tabular}
\end{table}

\subsubsection{Discussion and Lessons Learned}\label{sub:setup2lessons}
We start this section by discussing implementation issues of Experimental 
Setup 2, and then elaborate on the lessons learned from the results and 
evaluation, addressing both RQ-1 and RQ-2.

The main concern in this setup is to scale the implementation of the 
experiment from a corpus with a few hundred files to one with 67000 files. 
After all, we are interested in studying the behavior of the IR techniques and 
solutions on a realistic data set that mirrors the characteristics of the target 
problem. The application of IR models (VSM and LSA) is thereby constrained by 
two factors: memory usage and computational time. 

Since the term-by-document matrix is sparse, there exist efficient means to 
represent the matrix in memory, storing only non-zero values. However, this 
requires that matrix calculations support this format. Concretely, vectorized 
multiplication requires (by definition) 
that the multiplicands are stored as vectors. This is a requirement that we 
used in Step 3 (see Section~\ref{sub:setup2step3}) to our advantage. We created 
sub-matrices from the sparse term-by-document matrix that would fit in memory 
and distributed the cosine calculation for the VSM model among a cluster. 
Looking at the studies discussed in Section~\ref{sub:setup1lessons}, only a few 
applied IR models on large corpora. Table~\ref{tab:setup2related} compares the 
average corpus in this study with the largest corpora identified in the 
relevant literature. None of these studies analyzes the implications of the 
large corpora on memory consumption and applicability of the proposed 
approaches in an actual solution that could be used in industry. 

\begin{table}
 \caption{Large scale IR experiments on source code}\label{tab:setup2related}
 \centering
 \begin{tabular}{llll}
 \toprule
 \emph{Publications} & \emph{IR model} & \emph{\# Documents} & \emph{\# Unique 
 Terms} \\
 \midrule
 \cite{poshyvanyk_concept_2012} & LSA & 18,147 & 17,295 \\
 \cite{moreno_relationship_2013} & LSA/VSM & 34,375 & Not stated \\
 \cite{gay_use_2009} & VSM & 74,996 & Not stated \\
 \cite{poshyvanyk_combining_2007} & LSA & 86,208 & 56,863 \\
 \cite{liu_feature_2007} & LSA & 89,000 & 56,861 \\
 \midrule
 Experimental Setup 2 & VSM & 67,334 & 359,954 \\
 \bottomrule  
 \end{tabular}
\end{table}

%chain 1: 67332 documents / 347854 terms
%chain 3: 67334 documents / 359954 terms
%chain 4: 67331 documents / 341098 terms
%chain 5: 67331 documents / 341097 terms
%chain 9: 67332 documents / 347853 terms
%chain10: 67332 documents / 347857 terms

\begin{figure}
 \centering
 \includegraphics[scale=0.4]{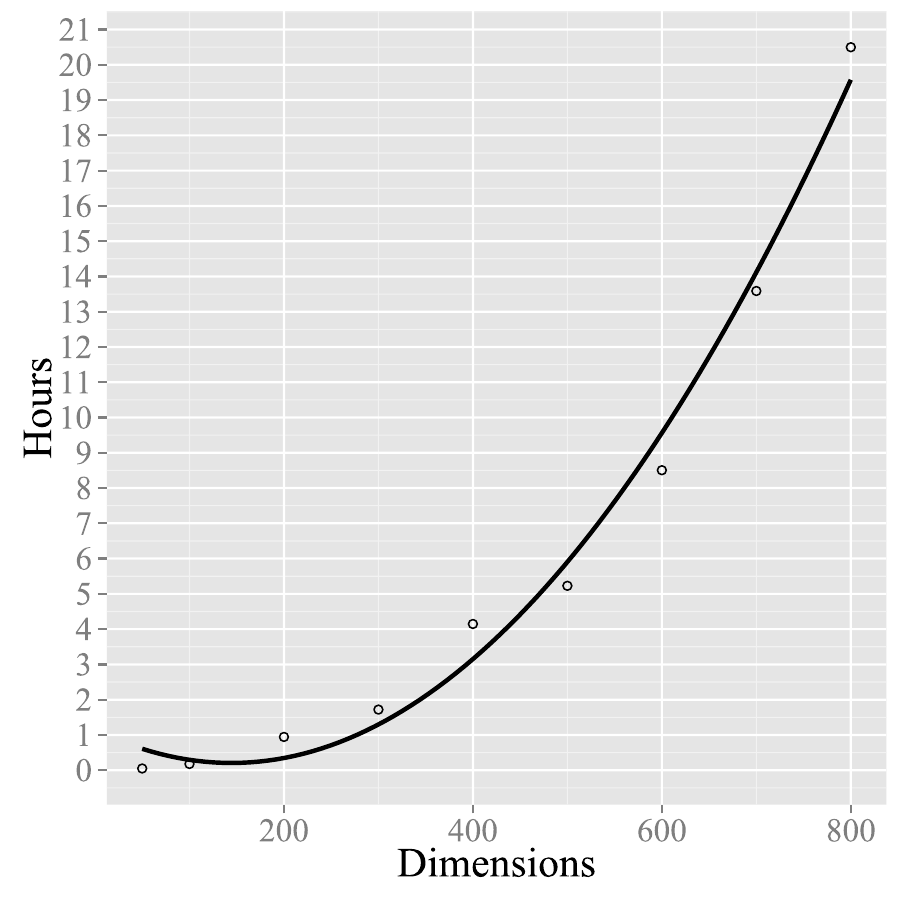}
 \caption{SVD runtime and dimensions}
 \label{fig:setup2svdruntime} 
\end{figure}

The second factor that constrains the applicability of the IR techniques is 
computational time. In Section~\ref{sub:setup2step3}, we have shown how the 
computational efficiency for the VSM model was increased by vectorization and 
parallelization of matrix multiplications. The LSA model requires a singular 
value decomposition of the term-by-document matrix. This operation is expensive 
with respect to execution time and difficult to parallelize for sparse 
matrices~\citep{kontoghiorghes_parallel_2006}. We ran a benchmark on a corpus 
with 35,387 documents and 193,861 terms, measuring the runtime of the SVD 
operation with various dimension reductions ($k={50, 100, 200, 300, 400, 500, 
600, 700, 800}$). We used the irlba 
package~\citep{baglama_irlba:_2014}, which allows partial SVD calculations in 
contrast to the standard SVD implementation in R. 
Figure~\ref{fig:setup2svdruntime} shows the measured runtime in hours versus 
the number of dimensions. For example, the runtime with $k=300$ amounts to 1 
hour and 43 minutes, whereas with $k=800$ the SVD computation requires 20 hours 
and 30 minutes. We fitted a simple quadratic linear regression model to the 
data, explaining 99\% of the observed variance (see 
Figure~\ref{fig:setup2svdruntime}), which we could use to extrapolate the 
runtime for higher values of $k$.

In Experimental Setup 1, we varied $k$ as a percentage of the number of 
documents in the corpus (80\%, 40\%, 20\%). If we would apply the same strategy 
in this setup with 67,000 documents, even with only a 20\% reduction 
($k=13700$), the SVD runtime would amount to 8,287 hours, or 345 days. This  
illustrates why determining an optimal $k$ is rather impractical and has been 
done only in a limited manner in the past (see discussion in 
Section~\ref{sub:setup1lessons}). However, experimentation represents the only 
way to determine the optimal $k$ for the data at 
hand~\citep{deerwester_indexing_1990}. Possible strategies that would make 
such experimentation possible, in particular with source code as documents, 
are to:
\begin{itemize}
 \item Reduce the number of terms in the corpus with fact extraction from C 
 source code files (Section~\ref{sub:setup2step2}). However, this approach 
 requires the implementation of fact extractors for the (potentially many) 
 programming languages occurring in the analyzed corpus.
 \item Exclude irrelevant documents from the corpus: we used every 
text-based document in the product repository as input. However, one could 
reduce this set to documents that are used in the product build 
process, effectively selecting only relevant documents as input. We  
explore this strategy in Experimental Setup 3 (Section~\ref{sub:setup3}).
 \item Use of parallel/distributed SVD implementations: we used a serial 
implementation which neither exploits multi-core processors nor can be run on a 
distributed system. Hence, one could explore solutions that parallelize the SVD 
computation, e.g. pbdR~\citep{ostrouchov_programming_2012}, 
SLEPc~\citep{hernandez_slepc:_2005} or gensim~\citep{rehurek_subspace_2011}. 
Note that incremental SVD computations, as suggested by \cite{brand_fast_2006} 
and \cite{jiang_incremental_2008}, would be of no benefit for model selection, 
since we are interested to vary $k$ and typically do not update the corpus.
\end{itemize}

Due to the inefficiency of the SVD computation, we decided to exclude LSA from 
our further experimentation, motivated by the instability of the VSM results in 
this setup and the uncertainty of the overall feasibility of the approach. It 
would be unwise to optimize the LSA model for computational efficiency 
when the approach, even with the simpler VSM model, turns out to be 
impractical. 

In Section~\ref{sub:setup2step4}, we argued that the VSM model seems to work on 
chains with certain characteristics, even though the Dsim measure does not 
reflect this. For example, the visual representation of the results of chain 6 
in Figure~\ref{subfig:chain10} indicates that there is a difference 
between the activated and deactivated corpus, even though Dsim (0,004, -0.009, 
-0.010) does not express this. Dsim discards too much information, i.e. it does 
not accurately represent the actual difference between a feature activated and 
deactivated corpus. Our first intuition on how to use IR techniques for test 
case selection (see discussion on solution development in 
Section~\ref{sub:idea}) would require to empirically identify a threshold value 
$\alpha$ for selecting/discarding a particular test case. Now however, after 
sampling more feature chains with a realistically sized document corpus, it 
seems  unlikely that we can determine a useful threshold value. Dsim clearly 
indicates that the difference between a feature activated and deactivated 
corpus cannot be measured by maximum similarity. Hence, a similarity threshold 
would not work either.

\begin{figure}
	\centering
	\includegraphics[scale=0.8]{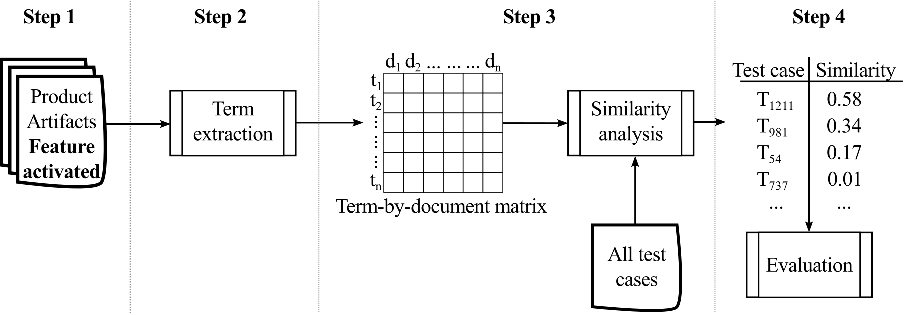}
	\caption{Ranking setup}
	\label{fig:ir_ranking} 
\end{figure}

We therefore decided to reshape the original solution description 
(Section~\ref{sub:idea}), allowing us to define an alternative evaluation 
metric to Dsim. Instead of using a test case as query and ranking product 
artifacts, we calculate an aggregated similarity of each test case to a product 
variant. This means that each test case is ranked with respect to a particular 
product variant. We can use this rank, in combination with the test case 
relevance information provided by a feature chain, to evaluate 1) the relative 
performance of IR techniques and 2) the test case selection performance.
This reformulation of the problem leads to a simplified experimental setup, as 
illustrated in Figure~\ref{fig:ir_ranking}. 
The main differences to the previous two setups (Figure~\ref{fig:ir}) are in 
Step 1 where we create only one feature activated corpus per chain and in Step 
4 where all existing system test cases are ranked according to their 
similarity to the feature activated corpus.

%http://irthoughts.wordpress.com/2008/02/13/lsi-how-many-dimensions-to-keep/

% What times do I need.
%Step 1
% - download, unpack of product
% - manual trace of feature to source code
%Step 2
% - normalizing text (lowercase, remove special chars and numbers), filter 
% stopwords, file filter, fact extraction
%Step 3
% - VSM calculation
% - LSI caluclation (on kernel only)

%SVD on 3.13.5 kernel
% 50 dim -> 175 sec / 1.4 GB
% 100 dim -> 639 Sec / 1.5 GB
% 200 dim -> 3398 / 1.8 GB
% 300 dim -> 6188 Sec / 2.2 GB RAM
% 400 dim -> 14926 Sec / 2.9 GB
% 500 dim -> 18822 Sec / 3.5 GB
% 600 dim -> 30619 Sec / 3.9 GB
% 700 dim -> 48913 Sec / 4.4 GB
% 800 dim -> 73795  Sec / 4.9 DB

% \begin{table}
%  \caption{Runtime/effort for each step in the 
% experiment}\label{tab:setup2effort}
%  \centering
%  \begin{tabular}{llll}
%  \toprule
%  Step 1 & Step 2 & Step 3 & Step 4 \\
%  \midrule
%  \midrule
%  \bottomrule  
%  \end{tabular}
% \end{table}

\subsection{Experimental Setup 3: Ranking}\label{sub:setup3}
In the previous setups we studied the feasibility and scalability of the 
approach, leading to refinements in individual steps, to the removal of the LSA 
model due to its computational cost from the experiment, and to a 
re-formulation of the problem in order to enable the evaluation. The objectives 
of this setup are therefore to:
\begin{itemize}
 \item adapt the experimental setup to the problem reformulation
 \item choose the correct statistical technique to evaluate and determine the 
factors that influence the IR technique performance
\end{itemize}

In this setup, we modified the configuration of the experiment in Steps 1 and 
4 (see Figure~\ref{fig:ir_ranking} and~\ref{fig:experimental_setups}). In Step 
4, we calculate the similarity of each test case (1600) to a product variant 
and then use this similarity score to rank the test-cases. Since we know which 
test cases are relevant for a product variant, we can then evaluate the IR 
techniques based on the respective test case rankings. 

Formally, we have in Experimental Setup 3 ten feature chains $I_{1\ldots10}$ 
that include ten features $F_{1\ldots10}$. We have ten feature vectors 
$A_{a1\ldots10}$ resulting in ten system variants 
$V(A_{a1\ldots10})$. To each feature, one or more test 
cases are mapped: in total, 65 test cases are mapped to ten features. We rank 
all 1600 test cases, which include the 65 mapped test cases, according to 
$sim(T_{1\ldots1600}, V(A_{a1\ldots10})))$. As for independent variables, we 
manipulate the content of test cases  $T_{1\ldots1600}$ and use two different 
summary statistics when calculating $sim$. 

The consequence of this problem reformulation, compared to Experimental Setup 1 
and 2, is an increase in computational cost: the similarity of each test case 
variant (4,800) to the product variant needs to be calculated (as opposed to 
the 3 test case variants in Experimental Setup 2). To make the ranking 
evaluation feasible, we employed two strategies to reduce the size of the 
document corpus and created two types of corpora:
\begin{enumerate}
 \item A minimal corpus that contains only the artifacts that are affected by a 
feature activation, reducing thereby noise stemming from artifacts that are 
common to all product variants. This allows us to pilot the new setup with a 
relatively short run-time. 
 \item An automatic corpus that represents the configured product as accurately 
as possible, reducing thereby the number of artifacts to what is actually 
composing a deployed product. 
\end{enumerate}

We call the first corpus ``minimal'' for two reasons: a) features are 
implemented in a small subset of the total files, hence the corpus is small 
compared to the second, ``automatic'' corpus; b) the difficulty level is 
minimized by removing noise caused by not relevant files. We call the second 
corpus ``automatic'' since we employ techniques to create this corpus without 
manual intervention, in contrast to the previous experimental setups.

\subsubsection{Step 1 - Corpus creation}\label{sub:setup3step1}
For the creation of the minimal corpus, we followed the same procedure as in 
Experimental Setup 2, i.e. manually tracing the impact of a feature 
activation to product artifacts. However, in this setup, we only included 
artifacts that were affected by a feature activation, thereby creating a 
minimal corpus. 

% Mapping of id's in this paper to chain id's:
% 1 -> 1
% 2 -> 4
% 3 -> 9
% 4 -> 3
% 5 -> 5
% 6 -> 10
% 7 -> 2
% 8 -> 6
% 9 -> 7
% 10 -> 8

\begin{table}
 \caption{Corpora sizes (\# of documents / \# of
terms)}\label{tab:setup3corpussize}
 \centering
 \begin{tabular}{llll}
 \toprule
 \emph{Chain} & \emph{Minimal corpus} & \emph{Automatic corpus} & \emph{Manual 
 corpus (setup 2)} \\
 \midrule
 1 & 10/1,668 & 11,078/68,619 & 67,332/347,854 \\
 2 & 3/260 & 11,078/68,613 & 67,331/341,098\\
 3 & 56/3,471 & 11,078/68,614 & 67,332/347,853\\
 4 & 2/150 & 11,078/68,613 & 67,334/359,954\\
 5 & 2/29 & 11,078/68,613 & 67,331/341,097\\
 6 & 29/1,638 & 11,078/68,611 & 67,332/347,857\\
 7 & 61/2,170 & 11,078/68,608 & N/A \\
 8 & 2/242 & 11,078/68,611 & N/A \\
 9 & 3/630 & 11,078/68,613 & N/A \\
 10 & 3/66 & 11,078/68,612 & N/A \\
 \bottomrule  
 \end{tabular}
\end{table}

The idea for creating an automatic corpus stems from the disadvantages 
of manual corpus creation, being inefficient (it must be repeated for every 
chain), error-prone, and most importantly, incomplete. With a manual 
configuration, only the traced option is considered and all other options are 
not implemented in the product artifacts. This means that the artifacts in a 
manually generated corpus do not correspond to the artifacts that compose a 
product that would be eventually installed and tested on a camera, leading to a 
larger, less accurate corpus.

For example, Listing~\ref{lis:makefile} shows an excerpt from a Makefile where 
a configuration option (line 2) determines whether 2 files (line 3) should be 
built or not. If \verb|CONFIG_A| is the traced option, a manual 
configuration would delete the source files corresponding to line 2. However, 
independently of whether \verb|CONFIG_B| in line 6 is activated, the 
corresponding files in line 7 would be included in a manually generated corpus. 

The same principle holds for preprocessor directives that realize configuration 
options. In Listing~\ref{lis:cpreprocessor}, assume \verb|CONFIG_C| 
to be the traced option (line 8). With a manually generated corpus, the 
configuration option in line 2 would not be evaluated, therefore including the 
file (line 3) into the corpus, independently whether the product is actually run 
on an ARM processor. This increases the size of the corpus, adds noise and does 
not reflect the product for which the system test cases where developed.

We addressed this issue by exploiting the product build system. The basic idea 
of our approach is to hook into the build process custom code that performs 
pre-processing operations on the files included in the product. As a result, we 
get the configured (preprocessed) source code and the compiled product as 
it is installed on the camera, including configuration files and documentation. 
In this way, we could create a corpus that corresponds to the tested product 
merely by configuring and building the product.

Table~\ref{tab:setup3corpussize} illustrates the size characteristics of the 
minimal and automatic corpus from this setup and the manual corpus from 
Experimental Setup 2. We reduced the average size for the automatic corpus by a
factor 6 compared to the manual corpus. The minimal corpus is significantly 
smaller, which allowed us to pilot the ranking approach.

% chain 1: 11078 documents / 68619 terms
% chain 2: 11078 documents / 68608 terms
% chain 3: 11078 documents / 68613 terms
% chain 4: 11078 documents / 68613 terms
% chain 5: 11078 documents / 68613 terms
% chain 6: 11078 documents / 68611 terms
% chain 7: 11078 documents / 68613 terms
% chain 8: 11078 documents / 68612 terms
% chain 9: 11078 documents / 68614 terms
% chain10: 11078 documents / 68611 terms

\lstset{language=[gnu]make, numbers=left, stepnumber=1, 
basicstyle=\footnotesize\ttfamily}
\begin{lstlisting}[frame=tb, caption={Makefile with conditional inclusion of 
files}, label=lis:makefile]
  [...]
  ifeq ($(CONFIG_A), y)
    ide-gd_mod-y += ide-disk.o ide-disk_ioctl.o
  endif
  [...]
  ifeq ($(CONFIG_B), y)
    ide-gd_mod-y += ide-floppy.o ide-floppy_ioctl.o
  endif
  [...]
\end{lstlisting}

\lstset{language=C}
\begin{lstlisting}[frame=tb, caption={Preprocessor directives with 
conditional inclusion of files and code}, label=lis:cpreprocessor]
 [...]
 #ifdef CONFIG_D
   #include <asm/irq.h>
 #endif
 [...]
 void led_classdev_unregister(struct led_classdev *led_cdev)
 {
   #ifdef CONFIG_C
     if (led_cdev->trigger)
       led_trigger_set(led_cdev, NULL);
   #endif
   [...]
 }
\end{lstlisting}

The second major difference to Experimental Setup 2 was to use all system test 
cases (1,600) and rank them instead of analyzing similarity between a product 
and 
the applicable test case(s). As test cases were stored in a database, we 
could easily automatize the construction of the 4,800 test case variants.

\subsubsection{Step 2 - Term extraction}\label{sub:setup3step2}
We applied the same process as in Experimental Setup 2.

\subsubsection{Step 3 - Similarity analysis}\label{sub:setup3step3}
In Experimental Setup 2, a similarity analysis for one product configuration 
and one test case variant required approximately one hour of computational 
time. 
With 4,800 test case variants, this would amount to a computation time of 200 
days per chain. However, with the reduced corpus sizes we could compute a chain 
within 2 (minimal corpus) respectively 17 (automatic corpus) hours, making the 
ranking approach feasible.

\subsubsection{Step 4 - Evaluation}\label{sub:setup3step4}
The evaluation procedure differs considerably from Experimental Setup 2 where 
we looked at the difference of a feature activated and deactivated corpus. In 
this setup, we have only a feature activated corpus and rank the test cases 
according to their similarity to that corpus. To create such a test case 
ranking, we compute a single similarity value using a summary statistic of 
the corpus' documents similarity to each test case. As summary statistics we 
chose maximum and mean similarity.

The independent variables in this setup are the test-case variants and the 
summary statistics. For the test case variants we have three levels (see 
Table~\ref{tab:iv}) while the summary statistics consist of two levels (maximum 
and mean). Since we have two independent variables and use the same data for 
all factors, we have a two-way repeated measures design. The dependent variable 
is the ranking position of a test-case, i.e. the lower the ranking position 
the better. In total, we have 65 ground truth instances, i.e. relevant test 
cases that are ranked between position 1 and 1600.

The evaluation goal in this setup is twofold: (1) we compare the effect of 
the independent variables on the ranking performance; (2) we study the 
ranking performance to understand whether it can support test case selection 
decisions.

\paragraph{(1) Comparison of ranking performance}
We formulate the following null hypotheses:
\begin{description}
\item[$H0_1$:] The test case content does not significantly affect the ranking 
performance of the VSM.
\item[$H0_2$:] The summary statistic does not significantly affect the ranking 
performance of the VSM.
\end{description}

Furthermore, we evaluate whether the size of the corpus impacts the 
ranking performance and formulate the third null hypothesis as follows:

\begin{description}
 \item[$H0_3$:] The corpus type does not significantly affect the 
ranking performance of the VSM.
\end{description}

\begin{table}
 \caption{Results of randomization tests (100,000 iterations) of hypotheses 
$H0_1$ and $H0_2$}\label{tab:setup3randomization}
 \centering
 \begin{tabular}{lll}
 \toprule
 \emph{Effect} & \emph{p-value minimal corpus} & \emph{p-value automatic 
 corpus} \\
 \midrule
 1. Test case content & 0.32022 & 0.11216 \\
 2. Summary statistic & 0.07580 & 0.98244 \\
 %Interaction of 1. and 2. & 0.50366 & 0.04903 \\
 \bottomrule  
 \end{tabular}
\end{table}

\begin{figure}
	\centering
	\subfloat[Minimal corpus]{\label{subfig:rankminimal}
		\includegraphics[scale=0.5]{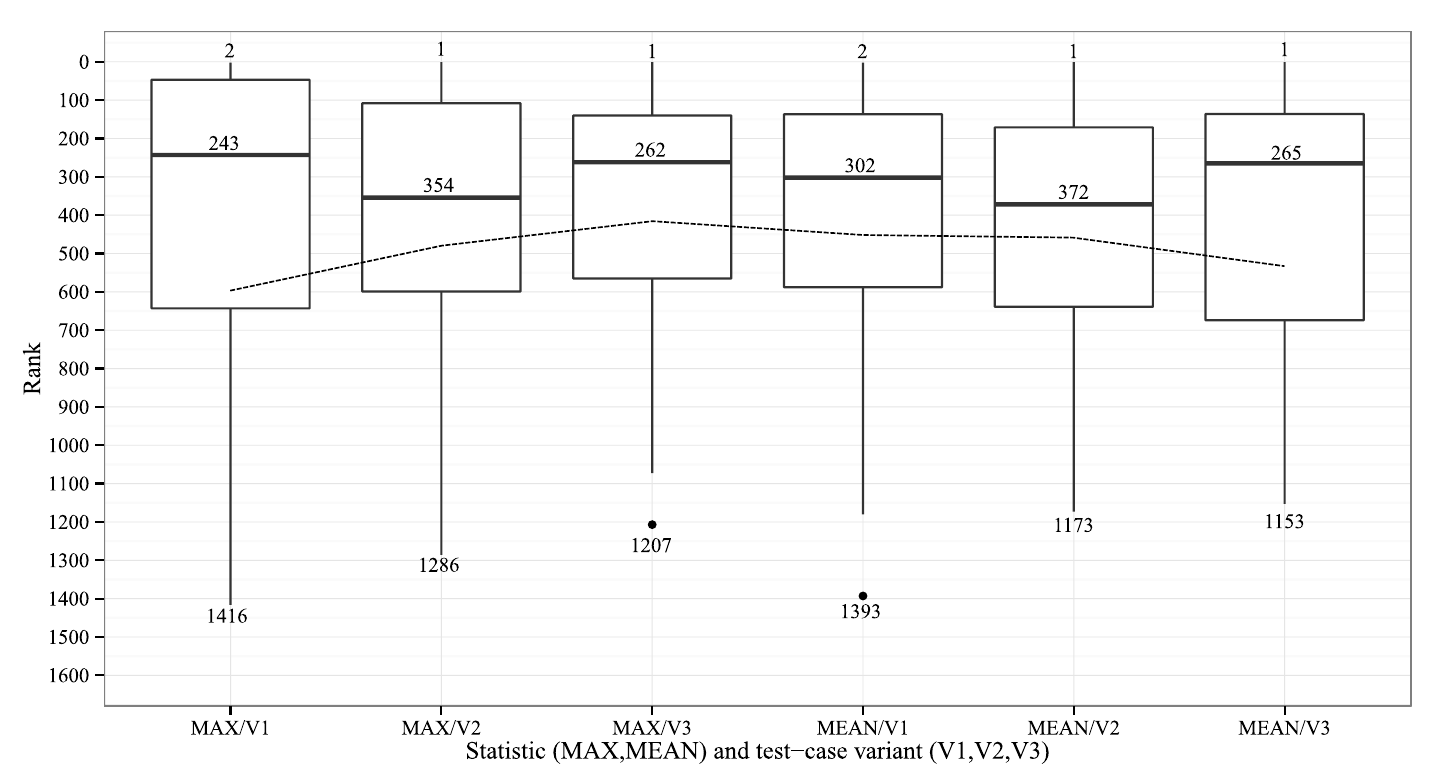}}
	\\
	\subfloat[Automatic corpus]{\label{subfig:rankautomatic}
		\includegraphics[scale=0.5]{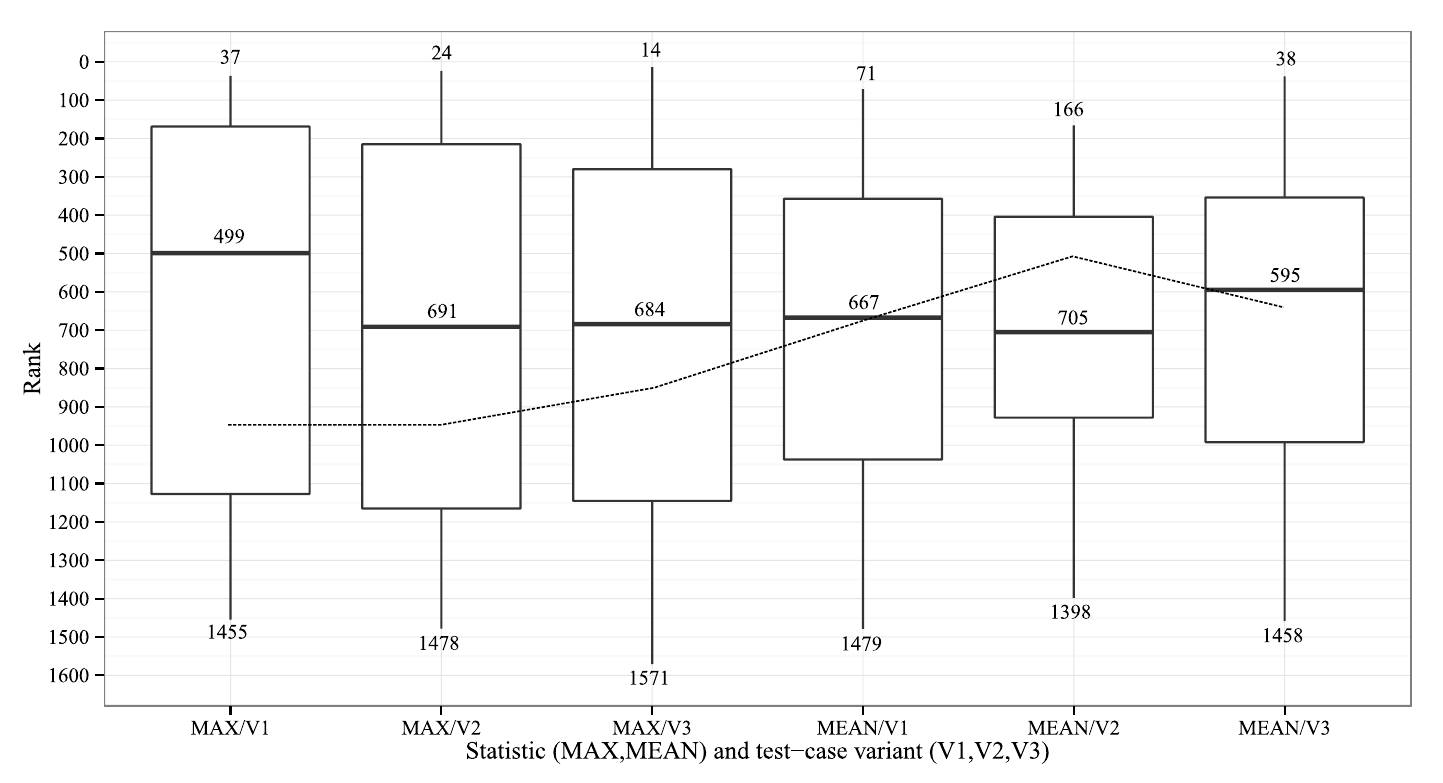}}
	\caption{Box plots of test case rankings - the dashed line indicates the 
		inter-quartile range differences}
	\label{fig:setup3rankresults} 
\end{figure}

Figure~\ref{fig:setup3rankresults} illustrates the ranking results in the form 
of box plots. The y-axis shows the ranking position (the numbers 
indicate minimum, median and maximum position) of the 65 test cases whereas the 
different box plots represent the two factors with two and three levels. Even 
though in both corpus types a large variance can be observed, the spread is 
more pronounced in the automatic corpus. This is an expected result as the 
minimal corpus contains less noise than the automatic corpus. Looking at the 
minimal corpus in Figure~\ref{subfig:rankminimal}, one can observe that, for 
the maximum statistic, the more specific the information in the test case the 
smaller is the variation (inter-quartile range, dashed line) in ranking results. 
This behavior is reversed for the mean statistic where the inter-quartile range 
increases with the specificity of the test case content. Looking at the 
automatic corpus in Figure~\ref{subfig:rankautomatic}, the impact of statistic 
and test case content on inter-quartile range is \emph{not} analogous to the 
one observed in the minimal corpus. This is an indication that the 
characteristics of the corpus (minimal vs. automatic) affect IR technique 
performance, not only by overall absolute values, but also in terms of the 
impact of the summary statistic and the test-case input on the ranking.

In order to test the stated hypotheses we used randomization 
tests~\citep{ludbrook_why_1998, edgington_randomization_2007} provided by the 
ez package~\citep{lawrence_ez:_2013} for R.
Table~\ref{tab:setup3randomization} lists the results of testing $H0_1$ and 
$H0_2$, which we consequently fail to reject at $\alpha=0.05$: neither test 
case content nor 
summary statistic significantly affect the test case ranking performance, both 
on the minimal and automatic corpus. However, we reject $H0_3$ with a 
p-value $<$ 0.001 (randomization test, 100,000 iterations).

\paragraph{(2) Test case selection decision support} 
Table~\ref{tab:setup3ranks} shows a subset of the results (for each 
chain only the test case with the highest rank is shown) in this 
experimental setup, both for the minimal and the automatic corpus. The first 
column indicates the chain and the number of test cases identified by the test 
engineers as being relevant for that feature. The remaining columns refer to 
the rank (out of 1,600) with the given test case variant and used summary 
statistic. For example, in the minimal corpus in Table~\ref{tab:setup3ranks}, 
the only test case in Chain 1 was ranked on position 81, using the complete 
test case and the mean as summary statistic. A lower rank indicates a better 
result since a relevant test case would appear higher in the ranked list of 
test cases.

Table~\ref{tab:setup3ranks} shows also the mean average precision 
(MAP)~\citep{liu_learning_2011} of the test case rankings. MAP allows us to 
interpret how useful the ranking is for the task at hand. MAP is calculated by 
taking the mean of the average precision (AP) over a set of queries. AP is 
calculated by averaging the precision at each relevant test case. For example, 
representing correct results as 1 and 0 otherwise, the AP for the query result 
$0~1~1~0~0~0~1$ is $(1/2 + 2/3 + 3/7)/3=0.53$. The MAP is then calculated by 
averaging the AP of a series of query results. A MAP of $0.1$ means that only 
every tenth ranked item is relevant, whereas with a MAP of $0.5$, every second 
item is relevant. 

\begin{center}
	\begin{threeparttable}
		\caption{Test case ranking positions and Mean Average 
			Precision}\label{tab:setup3ranks}
		\footnotesize
		\begin{tabular}{lrrrrrrr}
			\toprule
			\emph{Chain\tnote{1}} & \emph{MAX/V1} & \emph{MAX/V2} & 
			\emph{MAX/V3} & 
			\emph{MEAN/V1} & \emph{MEAN/V2} & \emph{MEAN/V3} \\
			\midrule
			\multicolumn{7}{c}{Minimal corpus - test case ranking position} \\
			\midrule
			\textbf{1 (1)} & 57 & 142 & 311 & 81 & 327 & 136 \\
			2 (2) & 1,416 & 1,286 & 955 & 1,393 & 1,166 & 992 \\
			\textbf{3 (5)} & 111 & 85 & 171 & 41 & 16 & 136 \\
			\textbf{4 (1)} & 2 & 1 & 4 & 2 & 1 & 4 \\
			5 (1) & 656 & 481 & 319 & 656 & 481 & 319 \\
			\textbf{6 (1)} & 37 & 107 & 105 & 4 & 28 & 125 \\
			\textbf{7 (34)} & 4 & 11 & 20 & 92 & 114 & 58 \\
			8 (10) & 891 & 766 & 660 & 912 & 795 & 670 \\
			9 (7) & 221 & 357 & 171 & 168 & 232 & 181 \\
			\textbf{10 (3)} & 29 & 9 & 17 & 20 & 5 & 11 \\
			\midrule
			MAP & 0.1140 & 0.1913 & 0.0879 & 0.1287 & 0.2032 & 0.0949 \\
			\midrule
			\multicolumn{7}{c}{Automatic corpus - test case ranking position} \\
			\midrule
			1 (1) & 461 & 524 & 165 & 497 & 630 & 1,365 \\
			2 (2) & 174 & 203 & 226 & 301 & 346 & 509 \\
			3 (5) & 67 & 185 & 14 & 229 & 194 & 631 \\
			4 (1) & 46 & 42 & 117 & 820 & 736 & 376 \\
			5 (1) & 888 & 805 & 1,112 & 1,074 & 793 & 821 \\
			6 (1) & 290 & 359 & 537 & 357 & 182 & 635 \\
			7 (34) & 892 & 1,120 & 1,229 & 71 & 225 & 38 \\
			8 (10) & 77 & 45 & 123 & 236 & 392 & 200 \\
			9 (7) & 37 & 66 & 109 & 477 & 450 & 347 \\
			10 (3) & 117 & 138 & 172 & 876 & 656 & 354 \\
			\midrule
			MAP & 0.0116 & 0.0122 & 0.0086 & 0.0058 & 0.0057 & 0.0062 \\
			\bottomrule  
		\end{tabular}
		\begin{tablenotes}
			\item [1] The number in parentheses indicates the total number of 
			test cases 
			that were mapped to that particular chain. 
		\end{tablenotes}
	\end{threeparttable}
\end{center}

In order to better understand the quality of the achieved rankings, we involved 
test engineers in the evaluation. Unfortunately, no test history was available 
for the studied product and feature chains, not allowing us to compare the 
achieved ranking with the selection from test engineers.
However, by querying test engineers we could establish whether test cases 
associated with a particular feature are relevant or not. For example, in Chain 
1, with the maximum summary statistic and test case variant 1, the known to be 
relevant test case was ranked on 
position 57 (see Table~\ref{tab:setup3ranks}, minimal corpus). This means that 
56 test cases were ranked above the relevant one. Our question to the test 
engineers was whether these higher ranked test cases were indeed relevant for 
the particular feature. We chose those chains for evaluation where the relevant 
test case was ranked below position 100. These six chains are indicated in bold 
typeface in Table~\ref{tab:setup3ranks}. Then, for each relevant test case 
(65), we selected those test cases which were ranked higher, or, in case the 
relevant test case was ranked at position 20 or lower, all top 20 test cases. 
This resulted in a set of 295 test cases, for each of which we identified the 
corresponding test area maintainers (eight in total). 
Table~\ref{tab:setup3precision} shows the true and false positive rates for 
each chain. According to the MAP in Table~\ref{tab:setup3ranks}, the IR 
technique with the mean summary statistic and variant 2 of the test case 
performed best. Hence, we chose that test case ranking to calculate the average 
precision, shown in Table~\ref{tab:setup3precision}. 

% Mapping of id's in this paper to chain id's:
% 1 -> 1
% 2 -> 4
% 3 -> 9
% 4 -> 3
% 5 -> 5
% 6 -> 10
% 7 -> 2
% 8 -> 6
% 9 -> 7
% 10 -> 8
\begin{table}
	\caption{Quality of the ranking results, judged by test are 
	maintainers}\label{tab:setup3precision}
	\centering
	\begin{tabular}{llllllll}
		\toprule
		\emph{Result / Chain} & 1 & 3 & 4 & 6 & 7 & 10 & \emph{Aggregate} \\
		\midrule
		\emph{Total test cases} & 48 & 29 & 25 & 19 & 141 & 33 & 295 \\
		\emph{True positives} & 9 & 12 & 7 & 4 & 48 & 21 & 101 \\
		\emph{False positives} & 39 & 17 & 18 & 15 & 93 & 12 & 194 \\
		\midrule
		\emph{Average Precision} & 0.21 & 0.61 & 0.42 & 0.24 & 0.10 & 0.70 &
		\textbf{0.38} \\
		\bottomrule  
	\end{tabular}
\end{table}

\subsubsection{Discussion and Lessons Learned}\label{sub:setup3lessons}
We start this section with a discussion of the results from Experimental Setup 
3, addressing RQ-2 and RQ-3, and then elaborate on the lessons learned from 
this evaluation, addressing RQ-1.

The first goal of this experimental setup was to study whether the test case 
content or summary statistic has an effect on the ranking performance of the 
VSM. The results (Table~\ref{tab:setup3randomization}) suggest that there is no 
statistically significant evidence for such an effect, even though there is a 
strong significant evidence that the ranking performance on minimal and 
automatic corpus differ (see the rejected $H0_3$). Looking at the results 
summarized in Table~\ref{tab:setup3randomization}, one can observe in the 
minimal corpus a tendency that the summary statistic affects the ranking 
performance. In the automatic corpus this tendency disappears. Even though this 
result is not statistically significant, it illustrates an important point: the 
characteristics of the data set on which the IR techniques are evaluated 
matter, even to the extent where the impact of a factor is reversed. 
Neglecting this can lead to sub-optimal IR technique configurations, e.g. when 
they are evaluated on a data set with characteristics that differ from the one 
where the technique is eventually applied for productive use. 
The experiments conducted by \cite{falessi_empirical_2013} illustrate the 
effect of data set difficulty on the performance of different IR techniques, 
confirming our observation that under different corpus characteristics (minimal 
vs. automatic), IR techniques behave differently.

The second goal of this experimental setup was to study to what extent the 
ranking provides test case selection decision support. Looking at 
Table~\ref{tab:setup3ranks}, the minimal corpus, we can observe that for Chains 
1, 3, 4, 6, 7 and 10 ranking positions below 100 were achieved. According to 
the MAP measure, the combination of V2 (test case except test 
steps) and the mean summary statistic would provide the best performance (i.e. 
every 5th test case is relevant). However, looking at the automatic corpus, the 
combination of V2 and the maximum summary statistic would provide the best 
performance. Furthermore, the MAP measure in the automatic corpus is by a 
factor 10 lower than in the minimal corpus, supporting the above conclusion 
that data set difficulty affects IR technique performance. We also evaluated 
how useful the best achieved rankings would be for test case selection. 
Table~\ref{tab:setup3precision} illustrates the achieved average precision with 
the selected best IR technique configuration. The results suggest that for 
some feature chains the approach works very well, e.g. for Chain 3 and 10 with 
a MAP $> 0.5$, which means that every second test case inspected by the test 
engineer was relevant. On the other hand, in Chain 7 only every 10th test case 
was assessed as relevant. Furthermore, we note that 
Table~\ref{tab:setup3precision} shows only the six out of ten feature chains 
for which we could achieve a ranking $< 100$ of one or more known to be 
relevant test cases. While these result seem underwhelming, they are comparable 
w.r.t. the achieved MAP of 0.38 to similar studies aiming at automatic 
traceability recovery. \cite{abadi_traceability_2008} traced code to 
documentation and experimented with 17 IR technique configurations, achieving a 
MAP between 0.04 and 0.66. \cite{cleary_empirical_2009} compared 6 concept 
location techniques, achieving a MAP between 0.006 and 0.06 (they attribute the 
low values to the large data set). \cite{qusef_recovering_2014} combine IR 
techniques with dynamic slicing to identify the classes tested by a test suite, 
achieving a MAP between 0.83 and 0.93. More recently, \cite{xia_dual_2015} 
associated bug reports with developers, achieving a MAP of 0.51. 

We turn now to the lessons learned from the statistical analysis. When we 
perform experiments comparing different IR techniques, we exercise them on the 
same set of tasks and determine the techniques' relative performance by testing 
whether their location parameter (e.g. mean or median) of some performance 
measure differs significantly~\citep{smucker_comparison_2007}. This general IR 
technique evaluation framework guides the selection and configuration of 
permissible statistical analyses, leading to the following considerations:

\begin{enumerate}
 \item \emph{Test assumptions}: The characteristics of the population 
distribution from which the dependent variables are sampled determine whether 
the assumptions (normality, independent samples, homogeneity of variance) of 
parametric tests are violated. An assumption that is likely to be violated is 
normality, caused by a skew to the right of the dependent variable. This 
is due to the inherent properties of some IR performance measures, e.g. a 
ranking cannot be negative.
 \item \emph{Repeated measures}: When comparing IR techniques we apply them on 
the same corpus, looking at the effect of different treatments (IR techniques) 
on the same subject (corpus). In other words, we have a paired (two IR 
techniques) or repeated (more than two IR techniques) measures design.
 \item \emph{Multiple testing}: When many different configurations of IR 
techniques are compared, we encounter the multiple testing 
problem~\citep{bender_adjusting_2001}: the more statistical tests are performed, 
the higher the likelihood that a true null hypothesis is rejected (false 
positive or Type I error). 
\end{enumerate}

Individually, the above considerations can be addressed by choosing 
statistical procedures fulfilling the given requirements. For example, if the 
assumptions of the parametric tests are violated, one can use non-parametric or 
distribution-free alternatives~\citep{sheskin_handbook_2000}. For a repeated 
measures design, paired difference tests or repeated measures ANOVA are viable 
choices~\citep{sheskin_handbook_2000}. The multiple testing problem can be 
addressed by applying adjustments to p-values or, depending on the particular 
study design, choosing test procedures that compensate for multiple 
comparisons~\citep{bender_adjusting_2001}.

In the context of evaluating IR techniques, we need to address all of the above 
stated considerations simultaneously and select a statistical procedure whose 
assumptions are not violated, allows for a repeated measures design and 
provides means to compensate for multiple testing. It turns out that these 
combined requirements are difficult to fulfill. If we assume that the normality 
assumption is violated, requiring a non-parametric test, and that we have a 
repeated measures design, we can only evaluate designs with at most one factor 
using Friedman's two-way analysis of variance by 
ranks~\citep{sheskin_handbook_2000}. This means that factorial designs to 
compare IR techniques can not be effectively evaluated with traditional 
statistical means, without allowing for violations of the test procedure and 
accepting a potential loss of statistical power. 

For example, \cite{biggers_configuring_2014} ignore potential normality 
violations (the dependent variable is a rank based on a similarity measure) and 
use regular ANOVA with five factors\footnote{Unfortunately it is not possible 
to determine whether this potential assumption violation had an impact on the 
outcome of the analysis since the raw data for this particular part (Part 1) of 
the case study has not been published by the authors.}. Subsequent analyses of 
interactions and main effects are performed with Kruskal-Wallis analysis of 
variance by ranks~\citep{sheskin_handbook_2000} (a non-parametric test) which 
indicates that the authors were aware of the potential violations of 
the parametric test assumptions. Further examples from recent journal 
publications illustrate the inherent difficulty to correctly evaluate IR 
techniques:
\begin{itemize}
 \item \cite{poshyvanyk_concept_2012} use Wilcoxon matched-pairs signed-ranks 
test~\citep{sheskin_handbook_2000}, acknowledging a potential violation of 
normality of their performance measure and taking advantage of the stronger 
statistical power of the paired test; however they do not compensate for 
multiple testing when evaluating the effect of stemming on the performance 
measure (48 tests).
 \item \cite{thomas_impact_2013} use Tukey's HSD test which is commonly 
recommended when all pair-wise comparisons in a data set need to be 
evaluated~\citep{sheskin_handbook_2000}. This test compensates for multiple 
tests, assumes however normality and homogeneity of variance. Even though 
the authors tested for the latter, normality is silently assumed. 
 \item \cite{falessi_empirical_2013} provide guidelines on the 
statistical evaluation of IR techniques, motivating the use of inferential 
statistics~\citep{sheskin_handbook_2000} by the need to ``check whether the 
observed difference could reasonable occur just by chance due to the random 
sample used for developing the model''~\citep{falessi_empirical_2013}. However, 
in the case-study presented in the same paper~\citep{falessi_empirical_2013} 
they refrain from using inferential statistical tests and rely on descriptive 
statistics to determine the best IR technique, not following their own 
best practice principles\footnote{They use inferential statistics to compare 
the best IR technique with the optimal combination of IR techniques, but this 
comparison is questionable since the preceding selection of best technique is
based on descriptive statistics.}.
\end{itemize}

We chose to follow the advise by \cite{smucker_comparison_2007} and used the 
randomization test~\citep{ludbrook_why_1998, edgington_randomization_2007, 
mittas_comparing_2008} to perform our IR technique comparison. The major 
advantage of randomization tests is that they do not assume any particular 
theoretical population distribution from which the sample is drawn but rather 
construct a sample distribution by permutations of the observed 
data~\citep{sheskin_handbook_2000}. This means that the skewness of our 
dependent variable (rank) is of no concern. A disadvantage of randomization 
tests is their computational cost: we performed the procedure with 100.000 
permutations on 65 observations on two factors (2x3 levels) in 11 hours. This 
is a considerable cost compared to generally instantaneous results of 
traditional statistical tests. 

Randomization tests are also applicable to repeated measure 
designs~\citep{sheskin_handbook_2000}. However, procedures to address the 
multiple testing problem in the context of repeated measure designs are 
difficult to implement, since comparisons occur between-subject factors, 
within-subject factors, or both~\citep{bender_adjusting_2001}. Hence, to the 
best of our knowledge, multiple testing is still an open issue, in the context 
of IR technique comparisons with more than two factors and under the 
constraints of a repeated measures design and potentially violated normality 
assumptions.

%variability mechanisms --> check research which works on different variants of 
%product (is there something out there??)

% Since test-case variants can be easily created, different combinations could 
% be tried out -> future work

\section{Conclusions and future work}\label{sec:conclusion}
This paper illustrated the experiences and lessons learned from developing and 
evaluating an approach for test case selection decision support based on 
Information Retrieval (IR) techniques. We rooted the solution development in 
the context of a large-scale industry experiment which allowed us to evaluate 
IR techniques in a realistic environment. The solution development was guided 
by incremental refinements of the experimental setup, both testing the 
scalability and performance of IR techniques. In order to provide insight for 
researchers, we reported on the design decisions, both in terms of experimental 
setup and IR technique implementation. We conclude by providing answers to the 
initially stated research questions.

\paragraph{RQ-1 To what extent can state-of-the-art IR techniques be applied 
in a large-scale industrial context?}
We identified a set of open questions that need to be addressed in order to 
bridge the gap between laboratory IR experiments and applications of IR in 
industry:

\begin{itemize}
 \item Comparative studies on IR techniques seldom optimize the parameters of 
all underlying IR models, leading to potentially unfair evaluations. This can 
even go as far as to reporting contradictory results, as we have shown in 
Section~\ref{sub:setup1lessons}, and has been also observed by 
\cite{grant_using_2013}. This poses a threat to both researchers and 
practitioners who might adopt IR techniques based on flawed evaluations. Future 
work is required on evaluating IR techniques both on standardized data sets but 
also on large-scale, industry-grade data.
 \item Parameter optimization is computationally expensive, particularly when 
applied on industry-grade, large data sets. Possible avenues for enabling 
parameter optimization efforts are: (1) investigating means to reduce the 
number of terms in a corpus, without affecting the IR models performance. Such 
techniques would however be very application and data specific, e.g. as the 
fact extraction and corpus-size reduction techniques illustrated in 
Sections~\ref{sub:setup2step1} and \ref{sub:setup3step1} respectively; (2) 
using parallel/distributed implementations of the algorithms, e.g. Singular 
Value Decomposition, that power IR techniques. A third alternative would be to 
compare optimized parametric with non-parametric IR models, such as
Hierarchical Dirichlet Processes~\citep{teh_hierarchical_2006}. If 
non-parametric models would show to be consistently equivalent or better than 
parametric models, they would be the favorable solution in an industrial 
environment since they would require less customization effort.
 \item Determining the superiority of one IR technique over another should 
be based upon inferential statistical 
techniques~\citep{falessi_empirical_2013}. However, as we have elaborated in 
Section~\ref{sub:setup3lessons}, developing and evaluating a valid statistical 
technique that allows simultaneously for non-normal dependent variables, a 
repeated measures design and multiple comparisons, in the context of IR 
evaluation, is still an open issue. 
\end{itemize}

\paragraph{RQ-2 To what extent do the used software development artifacts 
influence the performance of IR techniques?}
We have experimented with different variants of both test case content, i.e.  
information within the test case description, and product artifacts, in 
particular the size of the document corpus. We could not determine a 
statistical significant performance difference between test case variants. 
However, we observed that the size of the corpus influences the performance 
considerably (observed in Experimental Setup 2, and shown with statistical 
significance in Experimental Setup 3). This confirms the observations by 
\cite{falessi_empirical_2013} who evaluated IR techniques at varying
difficulty levels. As a consequence, this means that IR techniques need to be 
evaluated on realistic, large dataset that mirror the difficulty of datasets 
encountered in real industry applications.

\paragraph{RQ-3 To what extent can the studied IR techniques support test case 
selection?}
We have evaluated the test selection performance on ten feature chains. In four 
chains, the feature ``signal'' in the product artifacts was too weak to induce 
an actionable ranking. In the other six feature chains, we tasked test 
engineers to rate the test case ranking with respect to their relevance for the 
corresponding feature. The results in terms of precision (MAP $=0.38$) are 
comparable to what has been achieved in similar studies that aimed at 
automating trace recovery (e.g. 
\cite{abadi_traceability_2008,cleary_empirical_2009,qusef_recovering_2014, 
xia_dual_2015}) However, while our results reflect the state of art, 
the approach based on IR techniques is not yet reliable enough to automate the 
test case selection process in a realistic industry setting.

\bigskip
In future work, we plan to study the 
properties of the features chains where the differentiation between a feature 
activated and deactivated corpus failed. Knowing these properties and being 
able to detect them in the corpus would allow us to provide suggestions on how 
to improve source code documentation, e.g. by better naming conventions using a 
company-wide glossary. Furthermore, by including meta-data from the version 
control system in the corpus, such as commit comments connected to feature 
activation/deactivation code, we could improve the test case rankings.

\bibliographystyle{spbasic}
\bibliography{p5}

\begin{thebibliography}{109}
\providecommand{\natexlab}[1]{#1}
\providecommand{\url}[1]{{#1}}
\providecommand{\urlprefix}{URL }
\expandafter\ifx\csname urlstyle\endcsname\relax
  \providecommand{\doi}[1]{DOI~\discretionary{}{}{}#1}\else
  \providecommand{\doi}{DOI~\discretionary{}{}{}\begingroup
  \urlstyle{rm}\Url}\fi
\providecommand{\eprint}[2][]{\url{#2}}

\bibitem[{Abadi et~al(2008)Abadi, Nisenson, and
  Simionovici}]{abadi_traceability_2008}
Abadi A, Nisenson M, Simionovici Y (2008) A {Traceability} {Technique} for
  {Specifications}. In: Proceedings 16th {International} {Conference} on
  {Program} {Comprehension} ({ICPC}), IEEE, Amsterdam, The Netherlands, pp
  103--112

\bibitem[{Antoniol et~al(1999)Antoniol, Canfora, De~Lucia, and
  Merlo}]{antoniol_recovering_1999}
Antoniol G, Canfora G, De~Lucia A, Merlo E (1999) Recovering code to
  documentation links in {OO} systems. In: Proceedings 6th {Working}
  {Conference} on {Reverse} {Engineering} ({WCRE}), IEEE, Atlanta, USA, pp
  136--144

\bibitem[{Antoniol et~al(2000)Antoniol, Canfora, Casazza, and
  De~Lucia}]{antoniol_information_2000}
Antoniol G, Canfora G, Casazza G, De~Lucia A (2000) Information retrieval
  models for recovering traceability links between code and documentation. In:
  Proceedings {International} {Conference} on {Software} {Maintenance}
  ({ICSM}), IEEE, San Jose, USA, pp 40--49

\bibitem[{Antoniol et~al(2002)Antoniol, Canfora, Casazza, De~Lucia, and
  Merlo}]{antoniol_recovering_2002}
Antoniol G, Canfora G, Casazza G, De~Lucia A, Merlo E (2002) Recovering
  traceability links between code and documentation. IEEE Transactions on
  Software Engineering 28(10):970 -- 983

\bibitem[{Arcuri and Fraser(2011)}]{arcuri_parameter_2011}
Arcuri A, Fraser G (2011) On {Parameter} {Tuning} in {Search} {Based}
  {Software} {Engineering}. In: Proceedings 3rd {International} {Conference} on
  {Search} {Based} {Software} {Engineering} ({SSBSE}), Springer, Szeged,
  Hungary, pp 33--47

\bibitem[{Asuncion et~al(2010)Asuncion, Asuncion, and
  Taylor}]{asuncion_software_2010}
Asuncion H, Asuncion A, Taylor R (2010) Software traceability with topic
  modeling. In: Proceedings 32nd {International} {Conference} on {Software}
  {Engineering} ({ICSE}), IEEE, Cape Town, South Africa, pp 95--104

\bibitem[{Babar et~al(2010)Babar, {Lianping Chen}, and
  Shull}]{babar_managing_2010}
Babar MA, {Lianping Chen}, Shull F (2010) Managing {Variability} in {Software}
  {Product} {Lines}. IEEE Software 27(3):89--91, 94

\bibitem[{Baglama and Reichel(2014)}]{baglama_irlba:_2014}
Baglama J, Reichel L (2014) irlba: {Fast} partial {SVD} by implicitly-restarted
  {Lanczos} bidiagonalization.
  \urlprefix\url{http://cran.r-project.org/web/packages/irlba/index.html}

\bibitem[{Basili and Caldiera(1995)}]{basili_improve_1995}
Basili V, Caldiera G (1995) Improve software quality by reusing knowledge and
  experience. Sloan Management Review 37(1):55--64

\bibitem[{Bender and Lange(2001)}]{bender_adjusting_2001}
Bender R, Lange S (2001) Adjusting for multiple testing - when and how? Journal
  of Clinical Epidemiology 54(4):343--349

\bibitem[{Berry(2014)}]{berry_svdpackc_2014}
Berry MW (2014) {SVDPACKC}. \urlprefix\url{http://www.netlib.org/svdpack/}

\bibitem[{Berry et~al(2006)Berry, Mehzer, Philippe, and
  Sameh}]{kontoghiorghes_parallel_2006}
Berry MW, Mehzer D, Philippe B, Sameh A (2006) Parallel {Algorithms} for the
  {Singular} {Value} {Decomposition}. In: Handbook of {Parallel} {Computing}
  and {Statistics}, 1st edn, CRC Press

\bibitem[{Biggers et~al(2014)Biggers, Bocovich, Capshaw, Eddy, Etzkorn, and
  Kraft}]{biggers_configuring_2014}
Biggers LR, Bocovich C, Capshaw R, Eddy BP, Etzkorn LH, Kraft NA (2014)
  Configuring latent {Dirichlet} allocation based feature location. Empirical
  Software Engineering 19(3):465--500

\bibitem[{Biggerstaff et~al(1993)Biggerstaff, Mitbander, and
  Webster}]{biggerstaff_concept_1993}
Biggerstaff TJ, Mitbander BG, Webster D (1993) The {Concept} {Assignment}
  {Problem} in {Program} {Understanding}. In: Proceedings 15th {International}
  {Conference} on {Software} {Engineering} ({ICSE}), IEEE, Baltimore, USA, pp
  482--498

\bibitem[{Blei et~al(2003)Blei, Ng, and Jordan}]{blei_latent_2003}
Blei DM, Ng AY, Jordan MI (2003) Latent {Dirichlet} {Allocation}. Journal of
  Machine Learning Research 3:993--1022

\bibitem[{Borg et~al(2013)Borg, Runeson, and Ardö}]{borg_recovering_2013}
Borg M, Runeson P, Ardö A (2013) Recovering from a decade: a systematic
  mapping of information retrieval approaches to software traceability.
  Empirical Software Engineering pp 1--52, in Print

\bibitem[{Brand(2006)}]{brand_fast_2006}
Brand M (2006) Fast low-rank modifications of the thin singular value
  decomposition. Linear Algebra and its Applications 415(1):20--30

\bibitem[{Broy(2006)}]{broy_challenges_2006}
Broy M (2006) Challenges in {Automotive} {Software} {Engineering}. In:
  Proceedings 28th {International} {Conference} on {Software} {Engineering}
  ({ICSE}), ACM, Shanghai, China, pp 33--42

\bibitem[{Calcote(2010)}]{calcote_autotools:_2010}
Calcote J (2010) Autotools: {A} {Practitioner}'s {Guide} to {GNU} {Autoconf},
  {Automake}, and {Libtool}. No Starch Press

\bibitem[{Chen and Babar(2011)}]{chen_systematic_2011}
Chen L, Babar MA (2011) A systematic review of evaluation of variability
  management approaches in software product lines. Information and Software
  Technology 53(4):344--362

\bibitem[{Cleary et~al(2009)Cleary, Exton, Buckley, and
  English}]{cleary_empirical_2009}
Cleary B, Exton C, Buckley J, English M (2009) An empirical analysis of
  information retrieval based concept location techniques in software
  comprehension. Empirical Software Engineering 14(1):93--130

\bibitem[{Cleland-Huang et~al(2011)Cleland-Huang, Czauderna, Dekhtyar, Gotel,
  Hayes, Keenan, Leach, Maletic, Poshyvanyk, Shin, Zisman, Antoniol, Berenbach,
  Egyed, and Maeder}]{cleland-huang_grand_2011}
Cleland-Huang J, Czauderna A, Dekhtyar A, Gotel O, Hayes JH, Keenan E, Leach G,
  Maletic J, Poshyvanyk D, Shin Y, Zisman A, Antoniol G, Berenbach B, Egyed A,
  Maeder P (2011) Grand {Challenges}, {Benchmarks}, and {TraceLab}:
  {Developing} {Infrastructure} for the {Software} {Traceability} {Research}
  {Community}. In: Proceedings 6th {International} {Workshop} on {Traceability}
  in {Emerging} {Forms} of {Software} {Engineering} ({TEFSE}), ACM, Honolulu,
  USA, pp 17--23

\bibitem[{Clements and Northrop(2001)}]{clements_software_2001}
Clements P, Northrop L (2001) Software {Product} {Lines}: {Practices} and
  {Patterns}, 3rd edn. Addison-Wesley Professional, Boston

\bibitem[{Corazza et~al(2012)Corazza, Di~Martino, and
  Maggio}]{corazza_linsen:_2012}
Corazza A, Di~Martino S, Maggio V (2012) {LINSEN}: {An} efficient approach to
  split identifiers and expand abbreviations. In: Proceedings 28th
  {International} {Conference} on {Software} {Maintenance} ({ICSM}), IEEE,
  Trento, Italy, pp 233--242

\bibitem[{Crawley(2007)}]{crawley_r_2007}
Crawley MJ (2007) The {R} {Book}, 1st edn. John Wiley \& Sons

\bibitem[{Cullum and Willoughby(2002)}]{cullum_lanczos_2002}
Cullum JK, Willoughby RA (2002) Lanczos {Algorithms} for {Large} {Symmetric}
  {Eigenvalue} {Computations}: {Vol}. 1: {Theory}, 2nd edn. SIAM

\bibitem[{De~Lucia et~al(2006)De~Lucia, Fasano, Oliveto, and
  Tortora}]{de_lucia_can_2006}
De~Lucia A, Fasano F, Oliveto R, Tortora G (2006) Can {Information} {Retrieval}
  {Techniques} {Effectively} {Support} {Traceability} {Link} {Recovery}? In:
  Proceedings 14th {International} {Conference} on {Program} {Comprehension}
  ({ICPC}), IEEE, Athens, Greece, pp 307--316

\bibitem[{De~Lucia et~al(2007)De~Lucia, Fasano, Oliveto, and
  Tortora}]{de_lucia_recovering_2007}
De~Lucia A, Fasano F, Oliveto R, Tortora G (2007) Recovering traceability links
  in software artifact management systems using information retrieval methods.
  ACM Transactions Software Engineering Methodology 16(4)

\bibitem[{De~Lucia et~al(2009)De~Lucia, Oliveto, and
  Tortora}]{de_lucia_assessing_2009}
De~Lucia A, Oliveto R, Tortora G (2009) Assessing {IR}-based traceability
  recovery tools through controlled experiments. Empirical Software Engineering
  14(1):57--92

\bibitem[{De~Lucia et~al(2011)De~Lucia, Di~Penta, Oliveto, Panichella, and
  Panichella}]{de_lucia_improving_2011}
De~Lucia A, Di~Penta M, Oliveto R, Panichella A, Panichella S (2011) Improving
  {IR}-based {Traceability} {Recovery} {Using} {Smoothing} {Filters}. In:
  Proceedings 19th {International} {Conference} on {Program} {Comprehension}
  ({ICPC}), IEEE, Kingston, Canada, pp 21--30

\bibitem[{Deerwester et~al(1990)Deerwester, Dumais, Furnas, Landauer, and
  Harshman}]{deerwester_indexing_1990}
Deerwester S, Dumais ST, Furnas GW, Landauer TK, Harshman R (1990) Indexing by
  {Latent} {Semantic} {Analysis}. Journal of the American Society for
  Information Science 41(6):391--407

\bibitem[{Dit et~al(2011{\natexlab{a}})Dit, Guerrouj, Poshyvanyk, and
  Antoniol}]{dit_can_2011}
Dit B, Guerrouj L, Poshyvanyk D, Antoniol G (2011{\natexlab{a}}) Can {Better}
  {Identifier} {Splitting} {Techniques} {Help} {Feature} {Location}? In:
  Proceedings 19th {International} {Conference} on {Program} {Comprehension}
  ({ICPC}), IEEE, Kingston, Canada, pp 11--20

\bibitem[{Dit et~al(2011{\natexlab{b}})Dit, Revelle, Gethers, and
  Poshyvanyk}]{dit_feature_2011}
Dit B, Revelle M, Gethers M, Poshyvanyk D (2011{\natexlab{b}}) Feature location
  in source code: a taxonomy and survey. Journal of Software Maintenance and
  Evolution: Research and Practice 25(1):53--95

\bibitem[{Dit et~al(2013)Dit, Panichella, Moritz, Oliveto, Di~Penta,
  Poshyvanyk, and De~Lucia}]{dit_configuring_2013}
Dit B, Panichella A, Moritz E, Oliveto R, Di~Penta M, Poshyvanyk D, De~Lucia A
  (2013) Configuring topic models for software engineering tasks in {TraceLab}.
  In: Proceedings 7th {International} {Workshop} on {Traceability} in
  {Emerging} {Forms} of {Software} {Engineering} ({TEFSE}), IEEE, San
  Francisco, USA, pp 105--109

\bibitem[{Dit et~al(2014)Dit, Moritz, Linares-Vásquez, Poshyvanyk, and
  Cleland-Huang}]{dit_supporting_2014}
Dit B, Moritz E, Linares-Vásquez M, Poshyvanyk D, Cleland-Huang J (2014)
  Supporting and accelerating reproducible empirical research in software
  evolution and maintenance using {TraceLab} {Component} {Library}. Empirical
  Software Engineering pp 1--39

\bibitem[{Dumais(1992)}]{dumais_lsi_1992}
Dumais ST (1992) {LSI} meets {TREC}: a status report. In: {NIST} special
  publication, National Institute of Standards and Technology, pp 137--152

\bibitem[{Dyba et~al(2005)Dyba, Kitchenham, and
  Jorgensen}]{dyba_evidence-based_2005}
Dyba T, Kitchenham B, Jorgensen M (2005) Evidence-based software engineering
  for practitioners. IEEE Software 22(1):58--65

\bibitem[{Ebert and Jones(2009)}]{ebert_embedded_2009}
Ebert C, Jones C (2009) Embedded {Software}: {Facts}, {Figures}, and {Future}.
  Computer 42(4):42--52

\bibitem[{Edgington and Onghena(2007)}]{edgington_randomization_2007}
Edgington E, Onghena P (2007) Randomization {Tests}, 4th edn. Chapman and
  Hall/CRC, Boca Raton, FL

\bibitem[{Engström and Runeson(2011)}]{engstrom_software_2011}
Engström E, Runeson P (2011) Software product line testing - {A} systematic
  mapping study. Information and Software Technology 53(1):2--13

\bibitem[{Engström et~al(2010)Engström, Runeson, and
  Skoglund}]{engstrom_systematic_2010}
Engström E, Runeson P, Skoglund M (2010) A systematic review on regression
  test selection techniques. Information and Software Technology 52(1):14--30

\bibitem[{Enslen et~al(2009)Enslen, Hill, Pollock, and
  Vijay-Shanker}]{enslen_mining_2009}
Enslen E, Hill E, Pollock L, Vijay-Shanker K (2009) Mining source code to
  automatically split identifiers for software analysis. In: Proceedings 6th
  {International} {Working} {Conference} on {Mining} {Software} {Repositories}
  ({MSR}), IEEE, Vancouver, Canada, pp 71--80

\bibitem[{Falessi et~al(2013)Falessi, Cantone, and
  Canfora}]{falessi_empirical_2013}
Falessi D, Cantone G, Canfora G (2013) Empirical {Principles} and an
  {Industrial} {Case} {Study} in {Retrieving} {Equivalent} {Requirements} via
  {Natural} {Language} {Processing} {Techniques}. Transactions on Software
  Engineering 39(1):18--44

\bibitem[{Feinerer et~al(2008)Feinerer, Hornik, and Meyer}]{feinerer_text_2008}
Feinerer I, Hornik K, Meyer D (2008) Text {Mining} {Infrastructure} in {R}.
  Journal of Statistical Software 25(5):1--54

\bibitem[{Feldman(1979)}]{feldman_make_1979}
Feldman SI (1979) Make - a program for maintaining computer programs. Software:
  Practice and Experience 9(4):255--265

\bibitem[{Gay et~al(2009)Gay, Haiduc, Marcus, and Menzies}]{gay_use_2009}
Gay G, Haiduc S, Marcus A, Menzies T (2009) On the use of relevance feedback in
  {IR}-based concept location. In: Proceedings 28th {International}
  {Conference} on {Software} {Maintenance} ({ICSM}), IEEE, Edmonton, Canada, pp
  351--360

\bibitem[{Gethers et~al(2011)Gethers, Oliveto, Poshyvanyk, and
  De~Lucia}]{gethers_integrating_2011}
Gethers M, Oliveto R, Poshyvanyk D, De~Lucia A (2011) On integrating orthogonal
  information retrieval methods to improve traceability recovery. In:
  Proceedings 27th {International} {Conference} on {Software} {Maintenance}
  ({ICSM}), IEEE, Williamsburg, USA, pp 133--142

\bibitem[{Goldberg(1989)}]{goldberg_genetic_1989}
Goldberg DE (1989) Genetic algorithms in search, optimization, and machine
  learning, 1st edn. Addison Wesley, Boston, USA

\bibitem[{Gorschek et~al(2006)Gorschek, Wohlin, Carre, and
  Larsson}]{gorschek_model_2006}
Gorschek T, Wohlin C, Carre P, Larsson S (2006) A {Model} for {Technology}
  {Transfer} in {Practice}. IEEE Software 23(6):88--95

\bibitem[{Gotel and Finkelstein(1994)}]{gotel_analysis_1994}
Gotel O, Finkelstein CW (1994) An analysis of the requirements traceability
  problem. In: Proceedings 1st {International} {Conference} on {Requirements}
  {Engineering} ({RE}), IEEE, Colorado Springs, USA, pp 94--101

\bibitem[{Graaf et~al(2003)Graaf, Lormans, and Toetenel}]{graaf_embedded_2003}
Graaf B, Lormans M, Toetenel H (2003) Embedded software engineering: the state
  of the practice. IEEE Software 20(6):61--69

\bibitem[{Grant et~al(2013)Grant, Cordy, and Skillicorn}]{grant_using_2013}
Grant S, Cordy JR, Skillicorn DB (2013) Using heuristics to estimate an
  appropriate number of latent topics in source code analysis. Science of
  Computer Programming 78(9):1663--1678

\bibitem[{Grossman and Frieder(2004)}]{grossman_information_2004}
Grossman D, Frieder O (2004) Information {Retrieval} - {Algorithms} and
  {Heuristics}, The {Information} {Retrieval} {Series}, vol~15, 2nd edn.
  Springer, New York, USA

\bibitem[{Guerrouj et~al(2011)Guerrouj, Di~Penta, Antoniol, and
  Guéhéneuc}]{guerrouj_tidier:_2011}
Guerrouj L, Di~Penta M, Antoniol G, Guéhéneuc YG (2011) {TIDIER}: an
  identifier splitting approach using speech recognition techniques. Journal of
  Software Maintenance and Evolution: Research and Practice 25(6):575--599

\bibitem[{Hernandez et~al(2005)Hernandez, Roman, and
  Vidal}]{hernandez_slepc:_2005}
Hernandez V, Roman JE, Vidal V (2005) {SLEPc}: {A} {Scalable} and {Flexible}
  {Toolkit} for the {Solution} of {Eigenvalue} {Problems}. ACM Transactions on
  Mathematical Software 31(3):351--362

\bibitem[{Hill et~al(2008)Hill, Fry, Boyd, Sridhara, Novikova, Pollock, and
  Vijay-Shanker}]{hill_amap:_2008}
Hill E, Fry ZP, Boyd H, Sridhara G, Novikova Y, Pollock L, Vijay-Shanker K
  (2008) {AMAP}: automatically mining abbreviation expansions in programs to
  enhance software maintenance tools. In: Proceedings 5th {International}
  {Working} {Conference} on {Mining} {Software} {Repositories} ({MSR}), ACM,
  Leipzig, Germany, pp 79--88

\bibitem[{Hooker(1995)}]{hooker_testing_1995}
Hooker JN (1995) Testing heuristics: {We} have it all wrong. Journal of
  Heuristics 1(1):33--42

\bibitem[{Islam et~al(2012)Islam, Marchetto, Susi, and
  Scanniello}]{islam_multi-objective_2012}
Islam M, Marchetto A, Susi A, Scanniello G (2012) A {Multi}-{Objective}
  {Technique} to {Prioritize} {Test} {Cases} {Based} on {Latent} {Semantic}
  {Indexing}. In: Proceedings 16th {European} {Conference} on {Software}
  {Maintenance} and {Reengineering} ({CSMR}), IEEE, Szeged, Hungary, pp 21--30

\bibitem[{Ivarsson and Gorschek(2011)}]{ivarsson_method_2011}
Ivarsson M, Gorschek T (2011) A method for evaluating rigor and industrial
  relevance of technology evaluations. Empirical Software Engineering
  16(3):365--395

\bibitem[{Iverson(1980)}]{iverson_notation_1980}
Iverson KE (1980) Notation {As} a {Tool} of {Thought}. Communications of the
  ACM 23(8):444--465

\bibitem[{Jiang et~al(2008)Jiang, Nguyen, Chen, Jaygarl, and
  Chang}]{jiang_incremental_2008}
Jiang HY, Nguyen TN, Chen IX, Jaygarl H, Chang CK (2008) Incremental {Latent}
  {Semantic} {Indexing} for {Automatic} {Traceability} {Link} {Evolution}
  {Management}. In: Proceedings 23rd {International} {Conference} on
  {Automated} {Software} {Engineering} ({ASE}), IEEE, L'Aquila, Italy, pp
  59--68

\bibitem[{Knaus(2013)}]{knaus_snowfall:_2013}
Knaus J (2013) snowfall: {Easier} cluster computing (based on snow).
  \urlprefix\url{http://cran.r-project.org/web/packages/snowfall/index.html}

\bibitem[{Lavesson and Davidsson(2006)}]{lavesson_quantifying_2006}
Lavesson N, Davidsson P (2006) Quantifying the {Impact} of {Learning}
  {Algorithm} {Parameter} {Tuning}. In: Proceedings 21st {National}
  {Conference} on {Artificial} {Intelligence} ({AAAI}), AAAI Press, Boston,
  USA, pp 395--400

\bibitem[{Lawrence(2013)}]{lawrence_ez:_2013}
Lawrence MA (2013) ez: {Easy} analysis and visualization of factorial
  experiments.
  \urlprefix\url{http://cran.r-project.org/web/packages/ez/index.html}

\bibitem[{Lawrie and Binkley(2011)}]{lawrie_expanding_2011}
Lawrie D, Binkley D (2011) Expanding identifiers to normalize source code
  vocabulary. In: Proceedings 27th {IEEE} {International} {Conference} on
  {Software} {Maintenance} ({ICSM}), IEEE, Williamsburg, USA, pp 113--122

\bibitem[{Lee et~al(2012)Lee, Kang, and Lee}]{lee_survey_2012}
Lee J, Kang S, Lee D (2012) A {Survey} on {Software} {Product} {Line}
  {Testing}. In: Proceedings 16th {International} {Software} {Product} {Line}
  {Conference} ({SPLC}), ACM, Salvador, Brazil, pp 31--40

\bibitem[{Liu et~al(2007)Liu, Marcus, Poshyvanyk, and
  Rajlich}]{liu_feature_2007}
Liu D, Marcus A, Poshyvanyk D, Rajlich V (2007) Feature {Location} via
  {Information} {Retrieval} {Based} {Filtering} of a {Single} {Scenario}
  {Execution} {Trace}. In: Proceedings 22nd {International} {Conference} on
  {Automated} {Software} {Engineering} ({ASE}), ACM, Atlanta, USA, pp 234--243

\bibitem[{Liu(2011)}]{liu_learning_2011}
Liu TY (2011) Learning to {Rank} for {Information} {Retrieval}, 1st edn.
  Springer

\bibitem[{Lohar et~al(2013)Lohar, Amornborvornwong, Zisman, and
  Cleland-Huang}]{lohar_improving_2013}
Lohar S, Amornborvornwong S, Zisman A, Cleland-Huang J (2013) Improving {Trace}
  {Accuracy} {Through} {Data}-driven {Configuration} and {Composition} of
  {Tracing} {Features}. In: Proceedings 9th {Joint} {Meeting} on {Foundations}
  of {Software} {Engineering} ({FSE}), ACM, Saint Petersburg, Russia, pp
  378--388

\bibitem[{Lormans and van Deursen(2006)}]{lormans_can_2006}
Lormans M, van Deursen A (2006) Can {LSI} help reconstructing requirements
  traceability in design and test? In: Proceedings 10th {European} {Conference}
  on {Software} {Maintenance} and {Reengineering} ({CSMR}), IEEE, Bari, Italy,
  pp 47--56

\bibitem[{Ludbrook and Dudley(1998)}]{ludbrook_why_1998}
Ludbrook J, Dudley H (1998) Why {Permutation} {Tests} are {Superior} to t and
  {F} {Tests} in {Biomedical} {Research}. The American Statistician
  52(2):127--132

\bibitem[{Maletic and Marcus(2000)}]{maletic_using_2000}
Maletic J, Marcus A (2000) Using latent semantic analysis to identify
  similarities in source code to support program understanding. In: Proceedings
  12th {International} {Conference} on {Tools} with {Artificial} {Intelligence}
  ({ICTAI}), IEEE, Vancouver, Canada, pp 46--53

\bibitem[{Maletic and Marcus(2001)}]{maletic_supporting_2001}
Maletic J, Marcus A (2001) Supporting {Program} {Comprehension} {Using}
  {Semantic} and {Structural} {Information}. In: Proceedings 23rd
  {International} {Conference} on {Software} {Engineering} ({ICSE}), IEEE,
  Toronto, Canada, pp 103--112

\bibitem[{Maletic and Valluri(1999)}]{maletic_automatic_1999}
Maletic J, Valluri N (1999) Automatic software clustering via {Latent}
  {Semantic} {Analysis}. In: Proceedings 14th {International} {Conference} on
  {Automated} {Software} {Engineering} ({ASE}), IEEE, Cocoa Beach, USA, pp
  251--254

\bibitem[{Maletic et~al(2002)Maletic, Collard, and
  Marcus}]{maletic_source_2002}
Maletic J, Collard M, Marcus A (2002) Source code files as structured
  documents. In: Proceedings 10th {International} {Workshop} on {Program}
  {Comprehension} ({IWPC}), IEEE, Paris, France, pp 289--292

\bibitem[{Marcus and Maletic(2003)}]{marcus_recovering_2003}
Marcus A, Maletic J (2003) Recovering documentation-to-source-code traceability
  links using latent semantic indexing. In: Proceedings 25th {International}
  {Conference} on {Software} {Engineering} ({ICSE}), IEEE, Portland, USA, pp
  125 -- 135

\bibitem[{Marcus et~al(2004)Marcus, Sergeyev, Rajlich, and
  Maletic}]{marcus_information_2004}
Marcus A, Sergeyev A, Rajlich V, Maletic J (2004) An information retrieval
  approach to concept location in source code. In: Proceedings 11th {Working}
  {Conference} on {Reverse} {Engineering} ({WCRE}), IEEE, Delft, The
  Netherlands, pp 214--223

\bibitem[{Mittas and Angelis(2008)}]{mittas_comparing_2008}
Mittas N, Angelis L (2008) Comparing cost prediction models by resampling
  techniques. Journal of Systems and Software 81(5):616--632

\bibitem[{Moreno et~al(2013)Moreno, Bandara, Haiduc, and
  Marcus}]{moreno_relationship_2013}
Moreno L, Bandara W, Haiduc S, Marcus A (2013) On the {Relationship} between
  the {Vocabulary} of {Bug} {Reports} and {Source} {Code}. In: Proceedings 29th
  {International} {Conference} on {Software} {Maintenance} ({ICSM}), IEEE,
  Eindhoven, The Netherlands, pp 452--455

\bibitem[{Oliveto et~al(2010)Oliveto, Gethers, Poshyvanyk, and
  De~Lucia}]{oliveto_equivalence_2010}
Oliveto R, Gethers M, Poshyvanyk D, De~Lucia A (2010) On the {Equivalence} of
  {Information} {Retrieval} {Methods} for {Automated} {Traceability} {Link}
  {Recovery}. In: Proceedings 18th {International} {Conference} on {Program}
  {Comprehension} ({ICPC}), IEEE, Braga, Portugal, pp 68--71

\bibitem[{Ostrouchov et~al(2012)Ostrouchov, Chen, Schmidt, and
  Patel}]{ostrouchov_programming_2012}
Ostrouchov G, Chen WC, Schmidt D, Patel P (2012) Programming with {Big} {Data}
  in {R}. \urlprefix\url{http://r-pbd.org/}

\bibitem[{Panichella et~al(2013)Panichella, Dit, Oliveto, Di~Penta, Poshyvanyk,
  and De~Lucia}]{panichella_how_2013}
Panichella A, Dit B, Oliveto R, Di~Penta M, Poshyvanyk D, De~Lucia A (2013) How
  to {Effectively} {Use} {Topic} {Models} for {Software} {Engineering} {Tasks}?
  {An} {Approach} {Based} on {Genetic} {Algorithms}. In: Proceedings 35th
  {International} {Conference} on {Software} {Engineering} ({ICSE}), IEEE, San
  Francisco, USA, pp 522--531

\bibitem[{Perrouin et~al(2010)Perrouin, Sen, Klein, Baudry, and
  Le~Traon}]{perrouin_automated_2010}
Perrouin G, Sen S, Klein J, Baudry B, Le~Traon Y (2010) Automated and
  {Scalable} {T}-wise {Test} {Case} {Generation} {Strategies} for {Software}
  {Product} {Lines}. In: Proceedings 3rd {International} {Conference} on
  {Software} {Testing}, {Verification} and {Validation} ({ICST}), pp 459--468

\bibitem[{Pettersson et~al(2008)Pettersson, Ivarsson, Gorschek, and
  Öhman}]{pettersson_practitioners_2008}
Pettersson F, Ivarsson M, Gorschek T, Öhman P (2008) A practitioner's guide to
  light weight software process assessment and improvement planning. The
  Journal of Systems and Software 81(6):972--995

\bibitem[{Pohl and Metzger(2006)}]{pohl_software_2006}
Pohl K, Metzger A (2006) Software {Product} {Line} {Testing}. Communications of
  the ACM 49(12):78--81

\bibitem[{Poshyvanyk and Marcus(2007)}]{poshyvanyk_combining_2007}
Poshyvanyk D, Marcus A (2007) Combining {Formal} {Concept} {Analysis} with
  {Information} {Retrieval} for {Concept} {Location} in {Source} {Code}. In:
  Proceedings 15th {International} {Conference} on {Program} {Comprehension}
  ({ICPC}), IEEE, Banff, Canada, pp 37--48

\bibitem[{Poshyvanyk et~al(2006)Poshyvanyk, Gueheneuc, Marcus, Antoniol, and
  Rajlich}]{poshyvanyk_combining_2006}
Poshyvanyk D, Gueheneuc YG, Marcus A, Antoniol G, Rajlich V (2006) Combining
  {Probabilistic} {Ranking} and {Latent} {Semantic} {Indexing} for {Feature}
  {Identification}. In: Proceedings 14th {International} {Conference} on
  {Program} {Comprehension} ({ICPC}), IEEE, Athens, Greece, pp 137--148

\bibitem[{Poshyvanyk et~al(2012)Poshyvanyk, Gethers, and
  Marcus}]{poshyvanyk_concept_2012}
Poshyvanyk D, Gethers M, Marcus A (2012) Concept location using formal concept
  analysis and information retrieval. ACM Transactions on Software Engineering
  and Methodology 21(4)

\bibitem[{Qusef et~al(2014)Qusef, Bavota, Oliveto, De~Lucia, and
  Binkley}]{qusef_recovering_2014}
Qusef A, Bavota G, Oliveto R, De~Lucia A, Binkley D (2014) Recovering
  test-to-code traceability using slicing and textual analysis. Journal of
  Systems and Software 88:147--168

\bibitem[{Rothermel and Harrold(1996)}]{rothermel_analyzing_1996}
Rothermel G, Harrold M (1996) Analyzing regression test selection techniques.
  IEEE Transactions on Software Engineering 22(8):529--551

\bibitem[{Salton et~al(1975)Salton, Wong, and Yang}]{salton_vector_1975}
Salton G, Wong A, Yang CS (1975) A {Vector} {Space} {Model} for {Automatic}
  {Indexing}. Communications of the ACM 18(11):613--620

\bibitem[{Shepperd et~al(2014)Shepperd, Bowes, and
  Hall}]{shepperd_researcher_2014}
Shepperd M, Bowes D, Hall T (2014) Researcher {Bias}: {The} {Use} of {Machine}
  {Learning} in {Software} {Defect} {Prediction}. IEEE Transactions on Software
  Engineering 40(6):603--616

\bibitem[{Sheskin(2000)}]{sheskin_handbook_2000}
Sheskin DJ (2000) Handbook of {Parametric} and {Nonparametric} {Statistical}
  {Procedures}, {Second} {Edition}, 2nd edn. Chapman and Hall/CRC, Boca Raton

\bibitem[{Smucker et~al(2007)Smucker, Allan, and
  Carterette}]{smucker_comparison_2007}
Smucker MD, Allan J, Carterette B (2007) A comparison of statistical
  significance tests for information retrieval evaluation. In: Proceedings 16th
  {Conference} on {Information} and {Knowledge} {Management} ({CIKM}), ACM,
  Lisbon, Portugal, pp 623--632

\bibitem[{Steyvers and Griffiths(2007)}]{steyvers_probabilistic_2007}
Steyvers M, Griffiths T (2007) Probabilistic topic models. In: Handbook of
  {Latent} {Semantic} {Analysis}, 1st edn, Psychology Press

\bibitem[{Teh et~al(2006)Teh, Jordan, Beal, and Blei}]{teh_hierarchical_2006}
Teh YW, Jordan MI, Beal MJ, Blei DM (2006) Hierarchical {Dirichlet}
  {Processes}. Journal of the American Statistical Association
  101(476):1566--1581

\bibitem[{{The Apache Software
  Foundation}(2014)}]{the_apache_software_foundation_apache_2014}
{The Apache Software Foundation} (2014) Apache {Lucene} - {Apache} {Lucene}
  {Core}. \urlprefix\url{http://lucene.apache.org/core/}

\bibitem[{Thomas et~al(2013)Thomas, Nagappan, Blostein, and
  Hassan}]{thomas_impact_2013}
Thomas SW, Nagappan M, Blostein D, Hassan AE (2013) The {Impact} of
  {Classifier} {Configuration} and {Classifier} {Combination} on {Bug}
  {Localization}. IEEE Transactions on Software Engineering 39(10):1427--1443

\bibitem[{Thomas et~al(2014)Thomas, Hemmati, Hassan, and
  Blostein}]{thomas_static_2014}
Thomas SW, Hemmati H, Hassan AE, Blostein D (2014) Static test case
  prioritization using topic models. Empirical Software Engineering
  19(1):182--212

\bibitem[{Thörn(2010)}]{thorn_current_2010}
Thörn C (2010) Current state and potential of variability management practices
  in software-intensive {SMEs}: {Results} from a regional industrial survey.
  Information and Software Technology 52(4):411--421

\bibitem[{Unterkalmsteiner et~al(2014)Unterkalmsteiner, Gorschek, and
  Feldt}]{unterkalmsteiner_supplementary_2014}
Unterkalmsteiner M, Gorschek T, Feldt R (2014) Supplementary {Material} to
  "{Large}-scale {Information} {Retrieval} - an experience report from
  industrial application".
  \urlprefix\url{http://www.bth.se/com/mun.nsf/pages/autotcs-exp}

\bibitem[{Utting et~al(2012)Utting, Pretschner, and
  Legeard}]{utting_taxonomy_2012}
Utting M, Pretschner A, Legeard B (2012) A taxonomy of model-based testing
  approaches. Software Testing, Verification and Reliability 22(5):297--312

\bibitem[{Wohlin et~al(2000)Wohlin, Runeson, Höst, Ohlsson, Regnell, and
  Wesslén}]{wohlin_experimentation_2000}
Wohlin C, Runeson P, Höst M, Ohlsson MC, Regnell B, Wesslén A (2000)
  Experimentation in software engineering: an introduction. Kluwer Academic
  Publishers, Norwell

\bibitem[{Wohlin et~al(2012)Wohlin, Aurum, Angelis, Phillips, Dittrich,
  Gorschek, Grahn, Henningsson, Kagstrom, Low, Rovegard, Tomaszewski, van
  Toorn, and Winter}]{wohlin_success_2012}
Wohlin C, Aurum A, Angelis L, Phillips L, Dittrich Y, Gorschek T, Grahn H,
  Henningsson K, Kagstrom S, Low G, Rovegard P, Tomaszewski P, van Toorn C,
  Winter J (2012) The {Success} {Factors} {Powering} {Industry}-{Academia}
  {Collaboration}. IEEE Software 29(2):67--73

\bibitem[{Xia et~al(2015)Xia, Lo, Wang, and Zhou}]{xia_dual_2015}
Xia X, Lo D, Wang X, Zhou B (2015) Dual analysis for recommending developers to
  resolve bugs. Journal of Software: Evolution and Process In Print

\bibitem[{Yoo and Harman(2012)}]{yoo_regression_2012}
Yoo S, Harman M (2012) Regression testing minimization, selection and
  prioritization: a survey. Software Testing, Verification and Reliability
  22(2):67--120

\bibitem[{Zeimpekis and Gallopoulos(2006)}]{zeimpekis_tmg:_2006}
Zeimpekis D, Gallopoulos E (2006) {TMG}: {A} {MATLAB} {Toolbox} for
  {Generating} {Term}-{Document} {Matrices} from {Text} {Collections}. In:
  Grouping {Multidimensional} {Data}, Springer, Berlin, Heidelberg, pp 187--210

\bibitem[{Zhao et~al(2006)Zhao, Zhang, Liu, Sun, and Yang}]{zhao_sniafl:_2006}
Zhao W, Zhang L, Liu Y, Sun J, Yang F (2006) {SNIAFL}: {Towards} a static
  noninteractive approach to feature location. ACM Transactions on Software
  Engineering and Methodology 15(2):195--226

\bibitem[{Řehůřek(2011)}]{rehurek_subspace_2011}
Řehůřek R (2011) Subspace {Tracking} for {Latent} {Semantic} {Analysis}. In:
  Proceedings 33rd {European} {Conference} on {Advances} in {Information}
  {Retrieval} ({ECIR}), Springer, Dublin, Ireland, pp 289--300

\end{thebibliography}

\end{document}